\documentclass[12pt]{article} 

\usepackage{amsmath}
\usepackage{amssymb}
\usepackage{mathtools}
\mathtoolsset{showonlyrefs}

\usepackage[pdfencoding=auto]{hyperref}

\usepackage{tikz}
\usetikzlibrary{shapes}
\usetikzlibrary {positioning}

\usepackage{natbib}

\usepackage{algorithm}
\usepackage[noend]{algpseudocode}

\usepackage{array}
\newcolumntype{C}[1]{>{\centering\let\newline\\\arraybackslash\hspace{0pt}}m{#1}}

\usepackage{tabularx,ragged2e,booktabs,caption}

\usepackage{multirow,array}

\title{Inference on Extended-Spectrum Beta-Lactamase \emph{Escherichia coli} and \emph{Klebsiella pneumoniae} data through SMC$^2$}
\author{Lorenzo Rimella, Simon Alderton, Melodie Sammarro, Barry Rowlingson,\\ Derek Cocker, Nick Feasey, Paul Fearnhead and Christopher Jewell}

\begin{document}
 \maketitle


\begin{abstract}
    We propose a novel stochastic model for the spread of antimicrobial-resistant bacteria in a population, together with an efficient algorithm for fitting such a model to sample data. We introduce an individual-based model for the epidemic, with the state of the model determining which individuals are colonised by the bacteria. The transmission rate of the epidemic takes into account both individuals' locations, individuals' covariates, seasonality and environmental effects. The state of our model is only partially observed, with data consisting of test results from individuals from a sample of households taken roughly twice a week for 19 months. Fitting our model to data is challenging due to the large state space of our model. We develop an efficient SMC$^2$ algorithm to estimate parameters and compare models for the transmission rate. We implement this algorithm in a computationally efficient manner by using the scale invariance properties of the underlying epidemic model, which means we can define and fit our model for a population on the order of tens of thousands of individuals rather than millions. Our motivating application focuses on the dynamics of community-acquired Extended-Spectrum Beta-Lactamase-producing \emph{Escherichia coli} (\emph{E. coli}) and \emph{Klebsiella pneumoniae} (\emph{K. pneumoniae}), using data collected as part of the Drivers of Resistance in Uganda and Malawi project \citep{cocker2022drivers}. We infer the parameters of the model and learn key epidemic quantities such as the effective reproduction number, spatial distribution of prevalence, household cluster dynamics, and seasonality.
\end{abstract}

\section{Introduction}

Individual-based stochastic epidemic models offer a powerful approach to disentangling the complex nature of disease transmission in populations of interest, and have been shown to provide unprecedented insight into the determinants of risk in outbreak settings in humans, livestock, and plants \citep{probert2018real,jewell2009bayesian,deardon2010inference,vlek2013clustering,parry2014bayesian}.  Typically, these models comprise a state-transition process, where individuals transition between a discrete set of epidemiological states; for example the well-known SIR model assumes individuals start as \emph{susceptible} to infection, before progressing sequentially to \emph{infected}, and thereafter \emph{removed} (either recovered with solid immunity or dead). The ability to model the transition rates as a function of time, incorporating both the configuration of the states, individual-level covariates and known relationships between individuals, allows a detailed analysis of the importance of such features in a given outbreak setting.  

In general, inference for epidemic models is complicated by the need to account for censored event data (e.g. unobserved susceptible to infected transitions), or risk biased parameter estimates. For well-characterised medium-sized populations where all individuals are observed -- such as populations of farms, or patients within a hospital -- a Bayesian approach employing Markov chain Monte Carlo data augmentation (daMCMC) represents the state of the art \citep{jewell2009bayesian,deardon2010inference,vlek2013clustering}.  However, as the population and the number of censored transition events increase, or the fraction of the observable population decreases, these methods rapidly lose efficiency.  Moreover, for cyclic state-transition models in which individuals can experience more than one instance of any transition event, exploring the space of the number of transition events, as well as when they occurred, presents a severe implementational challenge.

A popular alternative is approximate Bayesian computation \citep{fearnhead2012constructing,sunnaaker2013approximate,kypraios2017tutorial} which requires only a simulator from the model to give samples from an approximation to the true posterior distribution. However, the quality of this approximation requires the specification of informative, low-dimensional summary statistics, and these can be difficult to construct \citep{barnes2012considerate, prangle2014semi}.

Another option is Particle MCMC (PMCMC) \citep{andrieu2010particle}, where the intractable likelihood is replaced by an estimate obtained using Sequential Monte Carlo (SMC) techniques. The appealing aspect of PMCMC methods is their exactness, in the sense that they are proven to target the true posterior distribution of the parameters. However, they are computationally expensive as they require to run an entire SMC for each MCMC step, and so it is unlikely to be computationally-practicable in individual-based epidemic models where the population size is large. PMCMC algorithms are not sequential as they use SMC only to estimate the likelihood. The recent innovation, SMC$^2$ \citep{chopin2013smc2}, is an SMC algorithm that allows parameter inference, only requiring PMCMC steps when we need to overcome particle degeneracy of the parameters. SMC$^2$ appears to have multiple appealing features for individual-based epidemic models: it is a sequential algorithm, it does not require too many PMCMC steps, and it provides an estimate of the marginal likelihood of the model.

In this paper, we apply the SMC$^2$ algorithm to an individual-level model of acquisition and loss of commensal antimicrobial resistance (AMR) carrying bacteria in a three study communities in Malawi.  As described in Section \ref{sec:data}, the study represents a typical scenario in which a cyclic stochastic state-transition model is desired to investigate the drivers of transmission, and the observed dataset represents a panel of individuals sampled sparsely from the population. We show the utility of SMC$^2$ for fitting a high-dimensional individual-based epidemic model like ours, identifying its advantages over other popular approaches for fitting such a model: it is easy to implement, it does not need any summary statistics, it is computationally feasible for large populations.

\subsection{Transmission of ESBL \emph{E. coli} and \emph{K. pneumoniae} in Malawi}\label{sec:data}

Our work is motivated by the challenge of fitting an individual based epidemic model for the spread of bacterial infection. The dataset consists of positive-negative sample results for colonisation with extended-spectrum $\beta$-lactamase (ESBL) producing \emph{Escherichia coli} (\emph{E. coli}) and \emph{Klebsiella pneumoniae} (\emph{K. pneumoniae}), individual ID, household ID, household location, individual-level variables: gender, income and age, extracted from the complete dataset \citep{cocker2022drivers}. The samples were collected in three study areas in Malawi: Chikwawa, Chileka and Ndirande, over a time span of about 1 year and 5 months (from 29-04-2019 to 24-09-2020) covering both the wet (November-April) and dry (May-October) seasons. Households involved in the study were sampled using an ``inhibitory with close pairs'' design extended to allow for sampling within sites with spatially heterogeneous populations \citep{chipeta2017inhibitory,cocker2022drivers}. The output of the collecting procedure is a time series with data appearing roughly twice a week (time sparsity) and some periods without samples (e.g. during the COVID outbreak). 


To analyse this data we introduce an individual-based epidemic model, where the state of the model determines which individuals are colonised on a given day. The dynamics of such models can be defined by specifying the rate that any colonised individual colonises an uncolonised individual and the rate at which a colonised individual recovers. As we are modelling anti-microbial resistant bacteria, a recovered individual is assumed to be susceptible to future colonisations. An individual-based model is flexible as it allows us to account for the different factors that affect the colonisation rate -- and we consider and estimate the effect of time-of-year, distance between individuals, whether individuals share the same home, and covariate information such as gender, income and age on the rate at which one individual infects another.

\begin{figure}[httb!]
    \centering
    \includegraphics[width=\textwidth]{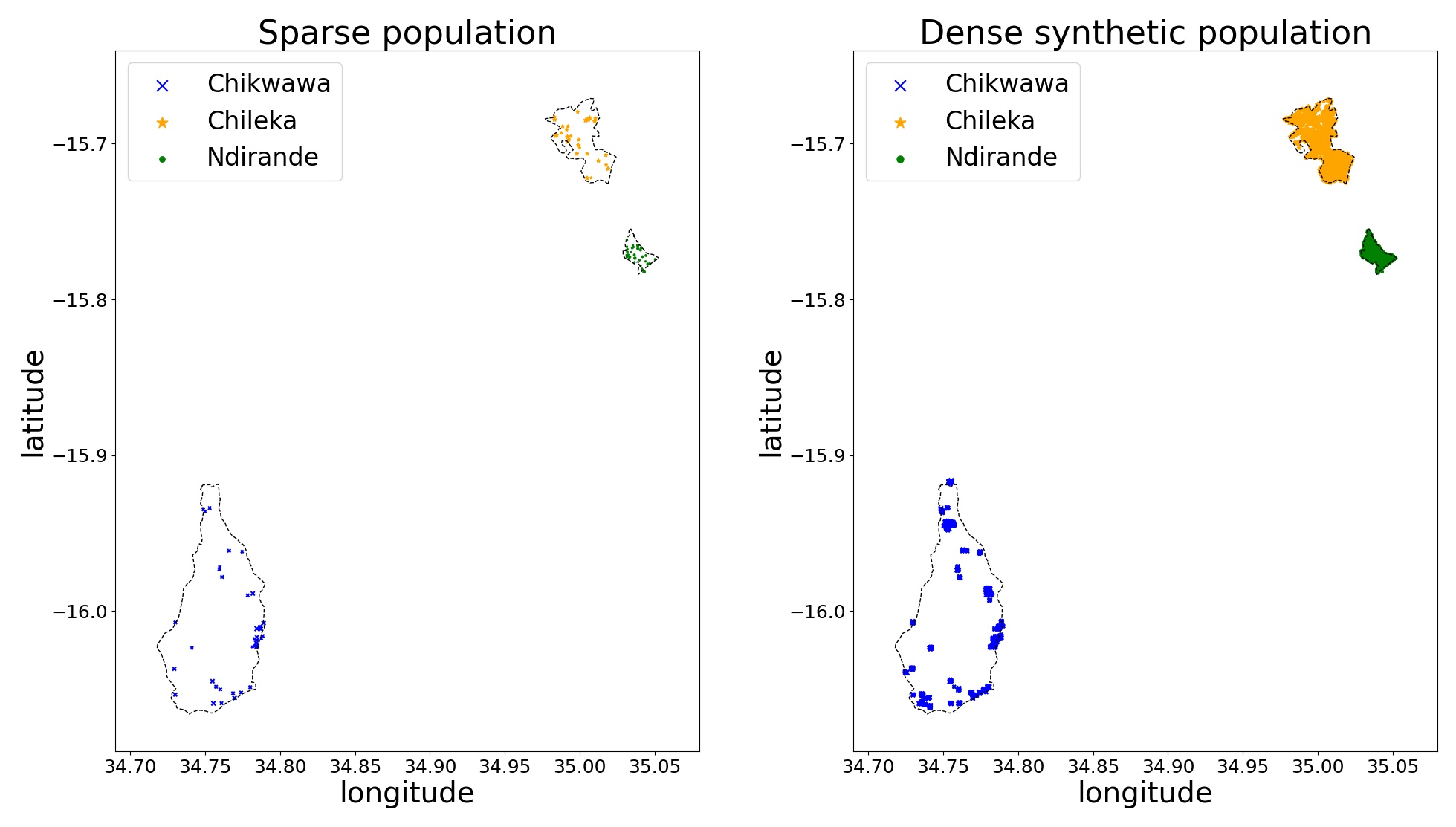}
    \caption{Households are represented with symbols whose size changes according to the household size. Sampled households are reported in the left plot (bottom left corner for Chikwawa, top right corner for Chileka and Ndirande). Synthetic households can be find in the right plot. Different symbols and different colors are associated with different areas.}
    \label{fig:synthetic_pop}
\end{figure}

To use such an approach we need the state of our model to include not only the colonisation status of the individuals that we sample but also the infection status of individuals in the population at large. Due to the scale-invariance of epidemic models, and in order to make inference computationally feasible we use a subsample of individuals from the population rather than all individuals and we checked that our results were robust to using such a sub-sample, by comparing results with different sub-sample sizes -- see the supplementary material. The samples of individuals were obtained by creating a synthetic population based on sampling households, but keeping all individuals within a household. We sampled household locations from the DRUM database household sample, the STRATAA census \citep{darton2017strataa}, OpenStreetMap (OSM) building data, or resampled from the DRUM database household sample itself with jitter, and individuals within each household were obtained from the DRUM database household sample with any missing member of the household sourced from the other households with the most similar characteristics. In total we generated a synthetic population of 36314 individuals for Ndirande distributed over 7949 households, 13337 individuals for Chileka distributed over 2888 households, and 9678 individuals for Chikwawa distributed over 2416 households. The population sizes are selected both to ensure good posterior estimates, see Section \ref{sec:Inference}, and to respect memory constraints on the GPU nodes of the HEC (High-End Computing) facility from Lancaster University. Figure \ref{fig:synthetic_pop} shows the data before and after the filling procedure. 

The synthetic population and the cleaning procedure are available at: \\ \href{https://github.com/LorenzoRimella/SMC2-ILM}{https://github.com/LorenzoRimella/SMC2-ILM}. 

More details on how to generate the synthetic population can be found at: \\ \href{https://zenodo.org/record/7007232#.Yv5EZS6SmUk}{https://zenodo.org/record/7007232\#.Yv5EZS6SmUk}.\\
The above link contains an anonymized version of the STRATAA dataset to avoid copyright issues. The anonymized version of the data provide the same synthetic population distribution. The authors can provide the real data, after authorization from the data owners, if needed. 

\section{Methodology}


\subsection{Agent-based UC model} \label{sec:UCmodel}
Consider a population size $n_I$, which varies according to the area (e.g. $n_I=36314$ in Ndirande), and define an index set $\{1,\dots,n_I\}$ with the notation $k \in \{1,\dots,n_I\}$ identifying uniquely an individual in the population. Let $C_t \in \{0,1\}^{n_I}$ be a vector representing the state of the population with respect to a single bacterium. For example, if we look at \emph{E. coli}, $C_t^{(k)}=1$ means the $k$th individual is colonised with \emph{E. coli} at time $t$, and $C_t^{(k)}=0$ means that they are uncolonised. Such a model is naturally defined in continuous time, however we consider an Euler discretisation to a discrete-time model. Initially we consider discretisatising to daily data, though more general discretisations are described in Section \ref{sec:Inference}. 

For a daily discretisation, we model $(C_t)_{t \geq 0}$ as a discrete time Markov chain, with one time unit corresponding to a day, where each component $k$ evolves as: 
\begin{equation}\label{eq:dynamic}
    C_0^{(k)} \sim \mathcal{B}(1-e^{-\lambda_0}), \quad C_{t+1}^{(k)} \sim 
    \begin{cases}
        \mathcal{B}\left(1-e^{-{\lambda}_t(\theta)^{(k)}}\right) & \text{if } C_{t}^{(k)}=0\\
        \mathcal{B}(e^{-\gamma})              & \text{if } C_{t}^{(k)}=1\\
    \end{cases} 
\end{equation}
where $\mathcal{B}(\cdot)$ is the Bernoulli random variable, $1-e^{-\lambda_0}$ is the initial probability of colonisation, ${\lambda}_t(\theta)^{(k)}$ is the transmission rate on individual $k$ at time $t$, with these depending on unknown parameters $\theta$, and $\gamma$ is the recovery rate, which is common across individuals. 

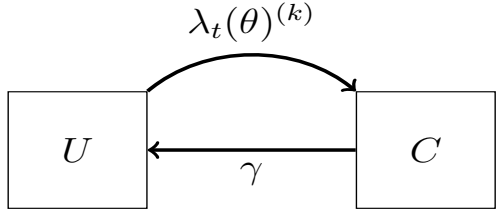
\begin{figure}[httb!]
	\centering
	\resizebox{0.5\textwidth}{0.15\textheight}{
	\begin{tikzpicture}[
	auto,
	node distance=60pt,
	box/.style={
		draw=black,
		anchor=center,
		align=center,
		minimum height=40pt,
		minimum width=40pt}
	]
	\node[box] (U) {$U$};
	\node[box, right=of U] (C) {$C$};
	
	\draw[->, very thick, looseness=1] (U) to node (UC) {$\lambda_t(\theta)^{(k)}$} (C);
	\draw[->, very thick] (C) to node (CU) {$\gamma$} (U);
	\end{tikzpicture}
	}
	\caption{A graphical representation of the UC model dynamic for a general individual $k$ described in equation \eqref{eq:dynamic}. $U$ stands for ``uncolonised'', while $C$ stands for ``colonised''.} \label{fig:UCmodel}
\end{figure}

The above construction considers the colonisation process of the bacteria as a Susceptible-Infected-Susceptible model (SIS) \citep{keeling2011modeling}, because the nature of the bacteria does not allow the individuals to become immune. We refer to this model as the Uncolonised-Colonised model, or the UC model for short. See Figure \ref{fig:UCmodel} for a graphical representation.

In \eqref{eq:dynamic}, the recovery rate $\gamma$ is assumed to be constant across individuals and over time, while the transmission rate ${\lambda}_t(\theta)^{(k)}$ is considered to be both time-varying and not homogeneous across individuals. We allow the transmission rate to take into account: within household transmission, between households transmission (and spatial distance), seasonality, the effect of individuals' covariates and a fixed effect from the environment. We define these effects separately and we then combine them to formulate ${\lambda}_t(\theta)^{(k)}$. 

Before listing the transmission rate components, we define the households as sets of individuals' indexes. Consider the household partition $\mathcal{H}$, which is a partition over the set $\{1,\dots,n_I\}$, then $\mathrm{H} \in \mathcal{H}$ stands for an household and $k \in \mathrm{H}$ is an individual inside the household $\mathrm{H}$. Throughout the manuscript, we use $\mathrm{H}^k$ to denote the household of individual $k$. 

Firstly, consider the within household transmission. We consider two possible models for the within household rate, each defined as:
\begin{equation}\label{eq:within_effect}
    {\lambda}^w_t(\beta_1)^{(k)} \coloneqq \beta_1 \frac{\sum_{k^\prime \in \mathrm{H}^k} C_t^{(k^\prime)} }{ \kappa_1(\mathrm{H}^k)},
\end{equation}
but a different choice of $\kappa_1(\mathrm{H}^k)$.
Here $\beta_1$ is a positive parameter and $\kappa_1(\mathrm{H}^k)$ is either the number of individuals in household $\mathrm{H}^k$, which we denote with $|\mathrm{H}^k|$, or $1$. These choices correspond respectively to a model in which colonisation rate is diluted by, or constant with respect to, increasing household size.

\begin{figure}[t]
    \centering
    \includegraphics[width=0.8\textwidth]{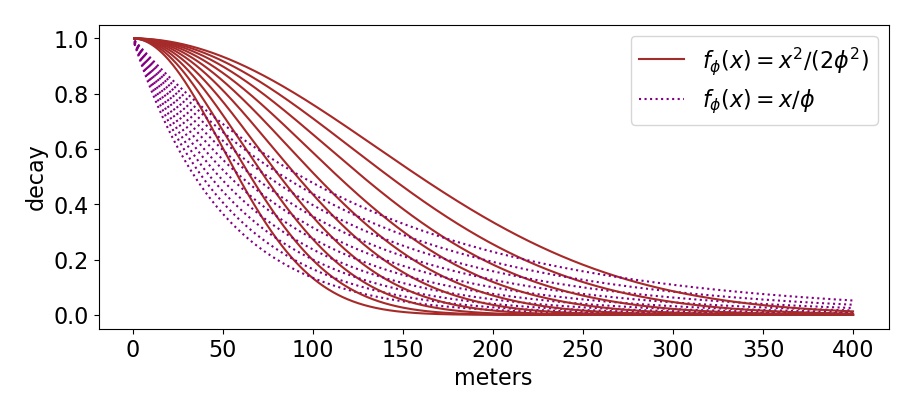}
    \caption{The considered decay functions, solid lines show the ``slow'' (Gaussian) decay for $\phi \in (e^{-3}, e^{-2})$, while dotted lines show the ``fast'' (exponential) decay for $\phi \in (e^{-3}, e^{-2})$. }
    \label{fig:decay_functions}
\end{figure}

Secondly, consider the between household transmission. We propose four models for the between household rate defined as:
\begin{equation}\label{eq:across_effect}
    {\lambda}^a_t(\beta_2, \phi)^{(k)} \coloneqq \beta_2 \sum_{\mathrm{H} \in \mathcal{H}} D_\phi^{\mathrm{H}^k,\mathrm{H}} \frac{\sum_{k^{\prime} \in \mathrm{H}} C_t^{(k^{\prime})} }{ \kappa_2(\mathrm{H})},
\end{equation}
but different choices of $\kappa_2(\mathrm{H})$ and $D_\phi^{\mathrm{H}^k,\mathrm{H}}$. Here $\beta_2,\phi$ are positive parameters and the model formulation varies according to $\kappa_2(\mathrm{H})$, which is either $1$ or $|\mathrm{H}|$, and $D_\phi^{\mathrm{H}^k,\mathrm{H}}$, which is a spatial kernel defined as:
\begin{equation}
    D_\phi^{\mathrm{H}^k,\mathrm{H}} \coloneqq 
    \begin{cases}
        e^{-f_{\phi}(d(\mathrm{H}^k, \mathrm{H}))} & \text{if } \mathrm{H}^k \neq \mathrm{H} \\
        0                        & \text{otherwise} 
    \end{cases},
\end{equation}
where $f_\phi(x)$ is either $x \slash \phi$ (exponential decay) or $x^2 \slash (2\phi^2)$ (Gaussian decay) and $d(\mathrm{H}^k, \mathrm{H})$ is the Euclidean distance between the rectangular coordinates of the households $\mathrm{H}^k$ and $\mathrm{H}$ in kilometers. $\kappa_2(\mathrm{H})$ has a similar interpretation to $\kappa_1(\mathrm{H})$. As already mentioned, $D_\phi^{\mathrm{H}^k,\mathrm{H}}$ is a spatial kernel which scales the transmission from each household with the distance, meaning that households that are far away from $\mathrm{H}^k$ are less likely to influence the colonisation process of $k$. In addition, $D_\phi^{\mathrm{H},\mathrm{H}}=0$ because the within household effect is modelled separately. This allows to decouple within household and between households transmissions and it improves identifiability. The form of $f_\phi$ distinguishes a fast ($f_\phi(x)=x \slash \phi$) from a slow ($f_\phi(x)=x^2 \slash (2\phi^2)$) decay at the origin, see Figure \ref{fig:decay_functions}. In practice, $f_\phi(x)=x^2 \slash (2\phi^2)$ implies a higher colonisation pressure from the neighbours.

Thirdly, we know that the prevalence of ESBL-producing \emph{E. coli} and \emph{K. pneumoniae} is higher during the wet season in Malawi \citep{lewis2021dynamics}, so we additionally define a seasonal effect:
\begin{equation}\label{eq:season_effect}
    s_t(\alpha) \coloneqq 1+ \alpha \cos(\text{frequency}\cdot t + \text{phase}), 
\end{equation}
where $\alpha$ is a parameter in $(0,1)$ and $\text{frequency}$, and $\text{phase}$ are chosen such that the peak of the function is in the middle of the wet season. A graphical representation is available in the supplementary material. Seasonality should not influence the within household transmission, because we expect the household environment to be stable over time. For this reason we use the seasonal effect as a multiplier of the between households transmission rate.

Next, the individuals' covariates might influence the transmission rate, hence we define an individual effect:
\begin{equation}\label{eq:individual_effect}
    I(\delta)^{(k)} \coloneqq e^{\langle (\text{covariates of k}), \delta \rangle},
\end{equation}
where $\delta$ is a 3-dimensional vector with each component referring to a different covariate (i.e. gender, income and age), ``$\text{covariates of k}$'' are the standardized covariates of individual $k$ and $\langle \cdot,\cdot \rangle$ denotes the scalar product between vectors. In contrast with the seasonal effect, we assume the individual effect to impact both the within household and the between households transmissions, hence we employ it as a global multiplier.

Finally, we also assume the presence of a fixed effect $\epsilon$, which is capturing the transmission that is not explained by the population dynamic and acts as a shift on the transmission rate.

The final formulation of the transmission rate combines these features, and is defined as:
\begin{equation}
    {\lambda}_t(\theta)^{(k)} = I(\delta)^{(k)} ({\lambda}^w_t(\beta_1)^{(k)} + s_t(\alpha){\lambda}^a_t(\beta_2, \phi)^{(k)})+\epsilon,
\end{equation}
where $\theta = (\beta_1, \beta_2, \phi, \alpha, \delta, \epsilon)$.
We have defined eight different combination of models, which vary according to $\kappa_1(|H|), \kappa_2(|H|), f_\phi$, these are further combine with setting or learning $\delta,\alpha,\epsilon$, for a total of forty-five models.

\subsection{Observation model}
We use $Y_t \in \{0,1, \text{NA} \}^{n_I}$ to indicate the test results of a specific bacterial species at time $t$, with $\text{NA}$ standing for ``not available'' (i.e. not tested at that time or not included in the study). For instance, if we look at ESBL \emph{K. pneumoniae} then $Y_t^{(k)}=0$ means that individual $k$ has tested negative for colonisation with ESBL \emph{K. pneumoniae} at time $t$ and reported in our dataset. We note that only a small subset of individuals is reported and it varies with time, hence we define the set $\mathrm{R}_t \subset \{1,\dots, n_I\}$ to represent the reported individuals at time $t$. Additionally, regarding the specificity and sensitivity of the test, even though $C_t^{(k)}=1$, there is a probability that we might get a false negative result (i.e. $Y_t^{(k)}=0$). Keeping these in mind we define the conditional distribution of $Y_t^{(k)}$ given $C_t^{(k)}$ as:
\begin{equation} \label{eq:emission}
    Y_t^{(k)}|C_t^{(k)} \sim 
    \begin{cases}
        \mathcal{B}(s_e)  & \text{if }k\in \mathrm{R}_t, \text{ } C_t^{(k)}=1\\ 
        \mathcal{B}(1-s_p) & \text{if }k\in \mathrm{R}_t, \text{ } C_t^{(k)}=0\\
        \text{NA} & \text{otherwise}
    \end{cases},  
\end{equation}
where $s_e,s_p$ are in $(0,1)$ and they represent the sensitivity and specificity of the test. As discussed in Section \ref{sec:data}, the data are sparse in both time and space. This sparsity is treated in \eqref{eq:emission} through the evolving set $\mathrm{R}_t$, which can be directly extracted from the data.

Sensitivity $s_e$, specificity $s_p$ along with the recovery rate $\gamma$, $\text{frequency}$ and $\text{phase}$ are treated as known. 

\section{Inference} \label{sec:Inference}
By definition $(C_t)_{t \geq 0}$ is an unobserved Markov chain and $Y_t$ is conditionally independent from all the other variables in the model given $C_t$, hence $(C_t, Y_t)_{t \geq 0}$ is a hidden Markov model  with finite state-space \citep{rabiner1986introduction, zucchini2009hidden}. Inference in a finite state-space hidden Markov model is naively pursued by computing the likelihood in closed form through the forward algorithm and then plugging it in a Markov chain Monte Carlo (MCMC) algorithm \citep{andrieu2003introduction, robert2004monte} to sample from the posterior distribution over the parameters of interest. However, in our case, this requires a marginalization over the latent state-space and so operations of the order $\mathcal{O}(2^{n_I})$, which is infeasible for even moderate size populations. 

We implement the  SMC$^2$ algorithm proposed by \cite{chopin2013smc2}, which sequentially target the posterior over both the parameters $\theta$ and the latent process $C_0,\dots, C_t$. 

\subsection{SMC and SMC$^2$} \label{subsec:smc_smc2}

SMC$^2$ can be intuitively seen as an SMC algorithm within an SMC algorithm, where the former controls the latent process $C_t$ and the latter guide the parameters $\theta$. The SMC algorithm for the latent process uses the Auxiliary Particle Filter (APF) \citep{pitt1999filtering,carpenter1999improved,johansen2008note} which proposes new states according to the distribution of $C_t|C_{t-1},Y_t$. 
In our model is simple to check that $C_t^{(k)}$ are conditionally independent given $C_{t-1},Y_t$. Furthermore the distribution of $C_t^{(k)}$ will differ depending on whether we have data on individual $k$ at time $t$. For individuals with data, the distribution of $C_t^{(k)}$ is $\mathcal{B}(p_t^{(k)})$ with: 
\begin{equation}\label{eq:p_t}
    p_t^{(k)} \coloneqq
        \begin{cases}
        \frac{ \left ( 1-e^{-\lambda_t(\theta)^{(k)}} \right ) \left [ s_e^{Y_t^{(k)}} \left ( 1-s_e \right )^{1-Y_t^{(k)}} \right]}{\left ( 1-e^{-\lambda_t(\theta)^{(k)}} \right ) \left [ s_e^{Y_t^{(k)}} \left ( 1-s_e \right )^{1-Y_t^{(k)}} \right] + e^{-\lambda_t(\theta)^{(k)}} \left [ s_p^{1-Y_t^{(k)}} \left ( 1-s_p \right )^{Y_t^{(k)}} \right]} & \text{if } C_{t-1}^{(k)}=0 \\
        \frac{e^{-\gamma} \left [ s_e^{Y_t^{(k)}} \left ( 1-s_e \right )^{1-Y_t^{(k)}} \right]}{e^{-\gamma} \left [ s_e^{Y_t^{(k)}} \left ( 1-s_e \right )^{1-Y_t^{(k)}} \right] + \left ( 1-e^{-\gamma} \right )\left [ s_p^{1-Y_t^{(k)}} \left ( 1-s_p \right )^{Y_t^{(k)}} \right]} & \text{if } C_{t-1}^{(k)}=1 \\
\end{cases}
\end{equation}
where the above is computed using \eqref{eq:dynamic} and \eqref{eq:emission} and with:
\begin{equation}\label{eq:p_0}
    p_0^{(k)} \coloneqq
        \frac{ \left ( 1-e^{-\lambda_0} \right ) \left [ s_e^{Y_0^{(k)}} \left ( 1-s_e \right )^{1-Y_0^{(k)}} \right]}{\left ( 1-e^{-\lambda_0} \right ) \left [ s_e^{Y_0^{(k)}} \left ( 1-s_e \right )^{1-Y_0^{(k)}} \right] + e^{-\lambda_0} \left [ s_p^{1-Y_0^{(k)}} \left ( 1-s_p \right )^{Y_0^{(k)}} \right]}.
\end{equation}
The APF for the UC-model is reported in Algorithm \ref{alg:SMC}, where a key role is played by the denominators in \eqref{eq:p_t}-\eqref{eq:p_0}:
\begin{equation}
    \begin{split}
        &w_0^{(k)} \coloneqq 
        \left ( 1-e^{-\lambda_0} \right ) \left [ s_e^{Y_0^{(k)}} \left ( 1-s_e \right )^{1-Y_0^{(k)}} \right] + e^{-\lambda_0} \left [ s_p^{1-Y_0^{(k)}} \left ( 1-s_p \right )^{Y_0^{(k)}} \right],\\
        &w_{t}^{(k)} \coloneqq
        \left \{ \left ( 1-e^{-\lambda_t(\theta)^{(k)}} \right ) \left [ s_e^{Y_t^{(k)}} \left ( 1-s_e \right )^{1-Y_t^{(k)}} \right] \right.\\
        & \quad \qquad \left.+ e^{-\lambda_t(\theta)^{(k)}} \left [ s_p^{1-Y_t^{(k)}} \left ( 1-s_p \right )^{Y_t^{(k)}} \right] \right \}\left (1-C_{t-1}^{(k)} \right )\\
        & \quad \qquad
        +e^{-\gamma} \left [ s_e^{Y_t^{(k)}} \left ( 1-s_e \right )^{1-Y_t^{(k)}} \right] + \left ( 1-e^{-\gamma} \right )\left [ s_p^{1-Y_t^{(k)}} \left ( 1-s_p \right )^{Y_t^{(k)}} \right]C_{t-1}^{(k)}.
    \end{split}
\end{equation}
If  $k \notin R_t$ (the set of sampled individuals) then $C_t^{(k)}|C_{t-1}^{(k)},Y_t$ is distributed as $C_t^{(k)}|C_{t-1}^{(k)}$ and follows \eqref{eq:dynamic}, with $w_t^{(k)}=1$.
Both $p_t^{(k)}$ and $w_t^{(k)}$ depend on $C_{t-1}^{(k)}$, and we make this dependence explicit in Algorithm \ref{alg:SMC} by writing $p_t^{p,(k)}$ and $w_t^{p,(k)}$, where $p$ is the particle index. Given the parameters $\theta$, the APF allows us to build particle approximations of the distribution of $C_t|Y_0,\dots,Y_t$ and estimates of the likelihood (i.e. the quantity $\mathcal{L}(\theta)$). 

\begin{algorithm}
	\caption{APF for UC-model}\label{alg:SMC}
	\begin{algorithmic}[1]
		\Require{$P$, $\theta$, $Y_1,\dots,Y_t$}
		\For{$p=1,\dots,P$}
		\For{$k=1,\dots,n_I$}
    		\State Compute $p_0^{(k)}$, sample $C_0^{p,(k)} \sim \mathcal{B}(p_0^{(k)})$ and compute $w_0^{(k)}$
		\EndFor
		\EndFor
		\State Set $w_0 \leftarrow \prod_{k=1}^{n_I} w_0^{(k)}$ and $\mathcal{L}_0(\theta) \leftarrow w_0$
		\For{$s = 1,\dots, t$}
		    \For{$p=1,\dots,P$}
		    \For{$k=1,\dots,n_I$}
			    \State Compute $p_s^{p,(k)}$, sample $C_s^{p,(k)} \sim \mathcal{B}(p_s^{p,(k)})$ and compute $w_s^{p,(k)}$
			\EndFor
			\State Set $w_s^p \leftarrow \prod_{k=1}^{n_I} w_s^{p,(k)}$
			\EndFor
			\State Set $\mathcal{L}_s(\theta) \leftarrow \mathcal{L}_{s-1}(\theta) \frac{1}{P}\sum_{p=1}^P w_s^p$ 
			\State Resample $C_s^{p}$ proportionally to $w_s^p$
		\EndFor
	\end{algorithmic}
\end{algorithm}

\begin{algorithm}
	\caption{SMC$^2$ for inference in UC-model}\label{alg:SMC2}
	\begin{algorithmic}[1]
		\Require{$P_{\theta}$,$P$, $\theta$, $Y_1,\dots,Y_t$}
		\For{$m=1,\dots,P_{\theta}$}
		\State Sample $\theta^m$ from the prior and set $w_{\theta}^m \leftarrow 1$
		\For{$p=1,\dots,P$}
		\For{$k=1,\dots,n_I$}
    		\State Compute $p_0^{(k)}$, sample $C_0^{m,p,(k)} \sim \mathcal{B}(p_0^{(k)})$ and compute $w_0^{(k)}$
		\EndFor
        \EndFor
        \State Compute $w_0 \leftarrow \prod_{k=1}^{n_I} w_0^{(k)}$, set $w_{\theta}^m \leftarrow w_0$ and $\mathcal{L}_0(\theta^m) \leftarrow w_0$
        \State Set the marginal likelihood $\mathcal{L}_0 \leftarrow w_0$
		\EndFor
		\For{$s = 1,\dots, t$}
            \For{$m=1,\dots,P_{\theta}$}
                \For{$p=1,\dots,P$}
                \For{$k=1,\dots,n_I$}
    			    \State Compute $p_s^{m,p,(k)}$ depending on $\theta^m$ and $C_{t-1}^{m,p}$
    			    \State Sample $C_s^{m,p,(k)} \sim \mathcal{B}(p_s^{m,p,(k)})$ and compute $w_s^{m,p,(k)}$
			\EndFor
			\State Set $w_s^{m,p} \leftarrow \prod_{k=1}^{n_I} w_s^{m,p,(k)}$ 
			\EndFor
			\State Set $w_{\theta}^m \leftarrow w_{\theta}^m \frac{1}{P}\sum_{p=1}^P w_s^{m,p}$ and $\mathcal{L}_s(\theta^m) \leftarrow \mathcal{L}_{s-1}(\theta^m) \frac{1}{P}\sum_{p=1}^P w_s^{m,p}$
			\State Set the marginal likelihood $\mathcal{L}_s \leftarrow \mathcal{L}_{s-1} \frac{1}{P_\theta P}\sum_{m=1}^{P_\theta}\sum_{p=1}^P w_\theta^m w_s^{m,p}$
			\EndFor
			\If{$w_{\theta}^m$ fulfill some degeneracy criteria}
			\State Resample $\theta^m$ proportionally to $w_{\theta}^m$
			\State Propose $\tilde{\theta}^m$ given $\theta^m$ and run Algorithm \ref{alg:SMC} up to $s$:
			\State \quad $\bullet$ get $\tilde{C}_s^{m,p}$ and $\mathcal{L}_s(\tilde{\theta}^{m})$
    	    \State Keep $({\theta^m}, {C}_s^{m,p}, \mathcal{L}_s({\theta}^{m}))$ or replace with $(\tilde{\theta}^m, \tilde{C}_s^{m,p}, \mathcal{L}_s(\tilde{\theta}^{m}))$
    	    \State Set $w_{\theta}^m \leftarrow 1$
			\EndIf
		\EndFor
	\end{algorithmic}
\end{algorithm}

Algorithm \ref{alg:SMC} requires us to know $\theta$, but it can be combined with another SMC algorithm to infer the parameters: resulting in the $\text{SMC}^2$ algorithm.
This algorithm stores at iteration $s$ a particle approximation to the joint posterior distribution of the parameters and latent state given the data up to the $s$th sample of households. These particle approximations are updated recursively from $s$ to $s+1$ by simulating the dynamics of the latent state between the associated time-points (particles for the latent process), weighting by the likelihood of the data at time $s+1$, and, if needed, resampling of the parameters (particles for the parameters). A key component of SMC$^2$ is the use of a particle MCMC step at resampling events, that allows for new parameter values to be sampled from their correct conditional distribution. An advantage of SMC$^2$ is that we can monitor the particle weights to get an estimate of the marginal likelihood for our model \citep{chopin2020introduction,chopin2013smc2}. Pseudocode for SMC$^2$ is reported in Algorithm \ref{alg:SMC2}, where from line 16 onward we briefly report the rejuvenation step and line 21 refers to a Metropolis-Hastings using the approximate likelihoods, more details are available in the supplementary material and in \cite{chopin2013smc2}.

To get an efficient implementation of SMC$^2$ we combine it with APF \citep{johansen2008note} and we simulate the new latent states over a time-step of up to seven days, see next paragraph. We also take advantage of the independence of our model over the two bacterial species and the three geographic regions so that we can parallelize the fitting procedure across species, regions and models for the infection rate.

\paragraph{\bf Time jumping APF within SMC$^2$}
The time sparsity of the data make the use of APF challenging for applications where $R_t = \emptyset$ for most $t$'s. Indeed, whenever $R_t = \emptyset$ we are sampling from the transition kernel in \eqref{eq:dynamic}, without correcting with observed data, which might lead to low effective sample size of the weights and high-variance estimates of the likelihood \citep{ju2021sequential,rimella2022approximating}. We propose to use a coarser time discretization, and simulate new individuals' state every $h$ days instead of every day. This can be done by generalizing \eqref{eq:dynamic}:
\begin{equation}\label{eq:dynamic_h}
    C_0^{(k)} \sim \mathcal{B}(1-e^{-\lambda_0}), \quad C_{t+h}^{(k)} \sim 
    \begin{cases}
        \mathcal{B}\left(1-e^{-{h\lambda}_t(\theta)^{(k)}}\right) & \text{if } C_{t}^{(k)}=0\\
        \mathcal{B}(e^{-h\gamma})              & \text{if } C_{t}^{(k)}=1\\
    \end{cases} 
\end{equation}
and by using this new dynamic to compute a $p_{t+h}^{(k)}$ as in \eqref{eq:p_t}, with $h\lambda_t(\theta)^{(k)}$ and $h\gamma$ appearing instead of $\lambda_t(\theta)^{(k)}$ and $\gamma$. Our main results are based on a weekly discretisation, so $h=7$. However, the DRUM data are not equally spaced in time, hence we define a simulation schedule between each pair of observation. The schedule is build by looking at a pair of times $t_1,t_2$ where $R_{t_1} \neq \emptyset$, $R_{t_2} \neq \emptyset$ and $R_t=\emptyset$ for all $t \in [t_1,t_2]$, and by dividing the interval $[t_1,t_2]$ in subintervals of size $7$ starting from $t_2$ and going backward (with the final being less than $7$ if $t_2-t_1$ is not divisible exactly). Note that choosing a bigger $h$ also affects the computational efficiency of the algorithm, by reducing the amount of simulation in the SMC$^2$ and so the computational cost. The validity of this procedure is checked empirically in Section \ref{sec:simulation}.

\begin{figure}[httb!]
     \centering
     \includegraphics[width=\textwidth]{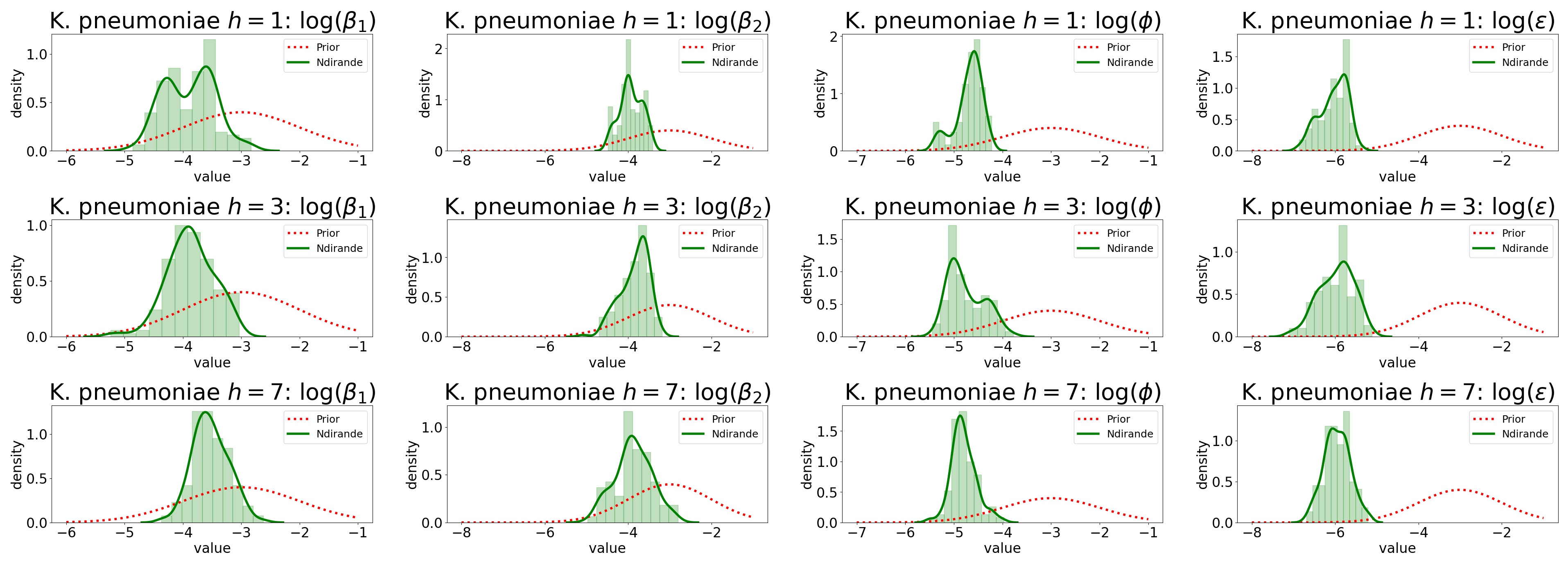}
     \caption{Posterior distribution for simulated data on Ndirande when $h=1,3,7$. On the columns from left to right: posterior distribution for $\beta_1,\beta_2,\phi,\epsilon$. On the rows from top to bottom: $h=1,3,7$.}
     \label{fig:invariance_time}
\end{figure}

\section{Simulation study} \label{sec:simulation}

As mentioned in the previous section, we run our experiments using an SMC$^2$ where the embedded APF is computing $C_{t+h}|C_t,Y_{t+h}$ by combining dynamic \eqref{eq:dynamic_h} with the emission distribution \eqref{eq:emission}. The advantages of such an approach are mainly computational and we also find it to improve inference (e.g. smoother posterior distributions, higher effective sample size). 

We test empirically the validity of this procedure on simulated data generated as follows:
\begin{itemize}
    \item we set $\log(\beta_1)=-2.8$, $\log(\beta_2)=-4.4$, $\log(\phi)=-4.6$, $\delta=(0,0,0)$, $\alpha=0.8$, $\log(\epsilon)=-6.1$, $\kappa_1(H)=\kappa_2(H)=|H|$, $f_{\phi}(x) = x \slash \phi$;
    \item we create a population as the one in Ndirande by merging the real data with the synthetic data;
    \item we simulate from \eqref{eq:dynamic} and report with \eqref{eq:emission}, using an $R_t$ as in the \emph{K. pneumoniae} data from DRUM.
\end{itemize}
The above data are then analysed with an SMC$^2$ algorithm where $h=1,3,7$. Each algorithm is run 4 times to check the reliability of the output. The run with the highest likelihood is then chosen and the corresponding posterior distributions are reported in Figure \ref{fig:invariance_time}. From Figure \ref{fig:invariance_time} we can notice that we are able to recover almost exactly $\beta_2,\phi$ and $\epsilon$, while $\beta_1$ is underestimated, which is due to the multimodality of the model and the sparsity of the observations. Increasing $h$ massively influence the computational cost which is around 316min for $h=1$, 101min for $h=3$ and 40min for $h=7$, and smooths the posterior distributions as shown in Figure \ref{fig:invariance_time}, but seems to introduce little bias. We also noticed that a higher $h$ is associated to a larger effective sample size of the parameters, indeed for $h=1$ we run $19$ rejuvenation steps, for $h=3$ we run $15$ rejuvenation steps and for $h=7$ we run $13$ rejuvenation steps.

\section{Analysis of DRUM data} \label{sec:results}

\subsection{Model selection}\label{sec:model_selection}

As already mentioned in the previous section, SMC$^2$ outputs both a sample from the posterior distribution over the parameters and a marginal likelihood estimate. We use the latter for model selection and the former to estimate the parameters and interpret the results. We perform inference in all the settings described in Section \ref{sec:UCmodel}, with $\lambda_0=0.13$, $\text{frequency}=2 \pi \slash 365.25$ to ensure a period of 1 year, $\text{phase}=0.55\pi$ to match wet-dry seasons in Malawi, recovery rate $\gamma= 1 \slash 10$ as suggested in \cite{lewEtAl19}, sensitivity $s_e=0.8$ and specificity $s_p=0.95$ \citep{cocker22}.

\begin{table}
    \caption{Table reporting the models with the highest posteriors under uniform priors. Model formulation changes according to $\kappa_1(|H|)$, $f_\phi(x)$, $\kappa_2(|H|)$, $\alpha$, $\delta$, $\epsilon$, but $\kappa_2(|H|)=|H|$, $\delta= (0,0,0)$ and $\epsilon$ with $\mathcal{N}(-3,1)$ are found to be the best. The best posterior score are colored in red.}\label{tab:posterior_table}
    \centering
    \fbox{
    \begin{tabular}{c|c|c|C{3cm}|C{3cm}}
         $\kappa_1(|H|)$ & $f_\phi(x)$ & $\alpha$ & marginal likelihood \emph{E. coli} & marginal likelihood \emph{K. pneu.}\\
         \hline
         $1$ & $x \slash \phi$ & $0.6$ &  \textcolor{red}{$0.861$} &
         $<0.005$\\
         \hline
        $|H|$ & $x^2 \slash (2\phi^2)$ & $0.4$ & $0.068$ & $0.088$\\
        \hline
         $|H|$ & $x^2 \slash (2\phi^2)$ & $0.8$ & $0.021$ &
         \textcolor{red}{$0.579$}\\
         \hline
         $|H|$ & $x \slash \phi$ & $0.6$ & $0.016$ &
         $0.093$\\
         \hline
         $|H|$ & $x^2 \slash (2\phi^2)$ & $0.2$ & $0.012$ &
         $0.015$\\
         \hline
         $|H|$ & $x \slash \phi$ & $0.2$ & $<0.005$ &
         $0.068$\\
         \hline
         $|H|$ & $x \slash \phi$ & $0.8$ & $<0.005$ &
         $0.078$
    \end{tabular}
    }
\end{table}

We run a total of 55 different models for each bacterial species--study area combination, with each SMC$^2$ run 4 times to ensure robustness of the output. For each bacteria and each model we compute the posterior distribution over the models under a uniform prior.  Finally, the marginal likelihoods are aggregated over the study areas, and we report the models with the 5 highest marginal likelihoods in Table \ref{tab:posterior_table}. We find that:
\begin{itemize}
    \item for \emph{E. coli} it is better to estimate $\epsilon$ rather than setting it to $0$, use an exponential decay in the spatial kernel rather than a Gaussian, set $\kappa_1(|H|)=1$ rather than $|H|$, set $\kappa_2(|H|)=|H|$ rather than $1$, set $\delta=(0,0,0)$ rather than estimate it and set the seasonality to $0.6$ rather than estimate it;
    \item for \emph{K. pneumoniae} it is better to estimate $\epsilon$ rather then setting it to 0, use a Gaussian decay in the spatial kernel rather than an exponential, set $\kappa_1(|H|)=|H|$ rather than $1$, set $\kappa_2(|H|)=|H|$ rather than $1$, set $\delta=(0,0,0)$ rather than estimate it and set the seasonality to $0.8$ rather than estimate it.
\end{itemize}
For both \emph{E. coli} and \emph{K. pneumoniae} we firstly try to learn $\delta$ and $\alpha$: for the former we find $\delta \approx (0,0,0)$ so we decide to set $\delta=(0,0,0)$; for the latter we find a posterior distribution over $\alpha$ in the interval $(0.4,0.6)$ for all the areas of study, see supplementary material, but this introduces a multimodal posterior distribution for the other parameters. We therefore take the pragmatic decision to learn $\alpha$ over the grid $(0.2,0.4,0.6,0.8)$ to improve identifiability. We note that setting $\delta$ and $\alpha$ also reduces the computational cost and gives higher marginal likelihood estimates.

\begin{figure}
    \centering
    \includegraphics[width=\textwidth]{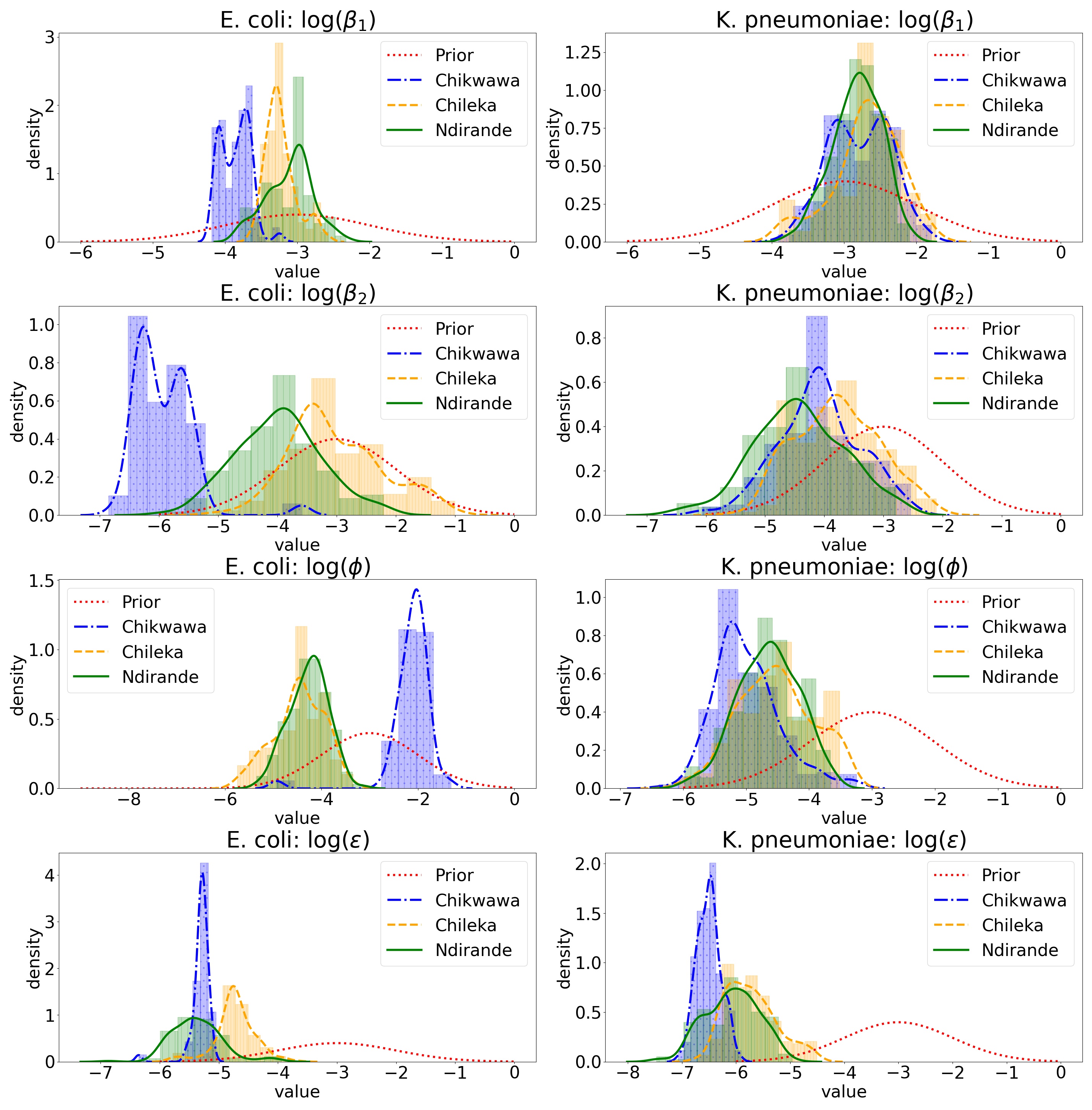}
    \caption{Estimated posterior distributions from the experiments' setting with the highest posteriors for each parameter, bacterial species and area of study. On the left column \emph{E. coli}, on the right column \emph{K. pneumoniae}. On the rows from top to bottom histograms and KDEs of $\beta_1, \beta_2, \phi, \epsilon$ in log scale. In each plot, blue is used for Chikwawa, orange for Chileka, green for Ndirande and red for the prior distribution over the parameter.}
    \label{fig:posterior_theta}
\end{figure}

\subsection{Parameter estimation}

The posterior distributions over the parameters $\beta_1,\beta_2,\phi,\epsilon$ from the model with highest marginal likelihood are reported in Figure \ref{fig:posterior_theta}, showing significant departure from their corresponding prior distributions. 

We notice that for \emph{E. coli}, $\kappa_1(|H|)=1$ suggesting a ``frequency dependent'' behaviour where the within household transmission increases with the number of colonised individuals in the household. However, we find that $\kappa_1(|H|)=|H|$ for \emph{K. pneumoniae} giving a ``density dependent'' behaviour of the force of colonisation with respect to the household size, i.e. a dilutional effect on the force of colonisation as the household size increases \citep{sammarroEtAl2022,Cocker2022}. For the between households transmission rate, we find $\kappa_2(|H|)=|H|$ for both bacteria, which is plausible since we are modelling contacts with colonised households. Indeed, considering the transmission rate on individual $k$, we can assume that once a contact between $k$ and household $H$ happens, the contact is going to be successful (resulting in the colonisation of $k$) according to the probability of meeting a colonised individual in $H$, which is the proportion of colonised in $H$. 

\begin{figure}[httb!]
    \centering
    \includegraphics[width=\textwidth]{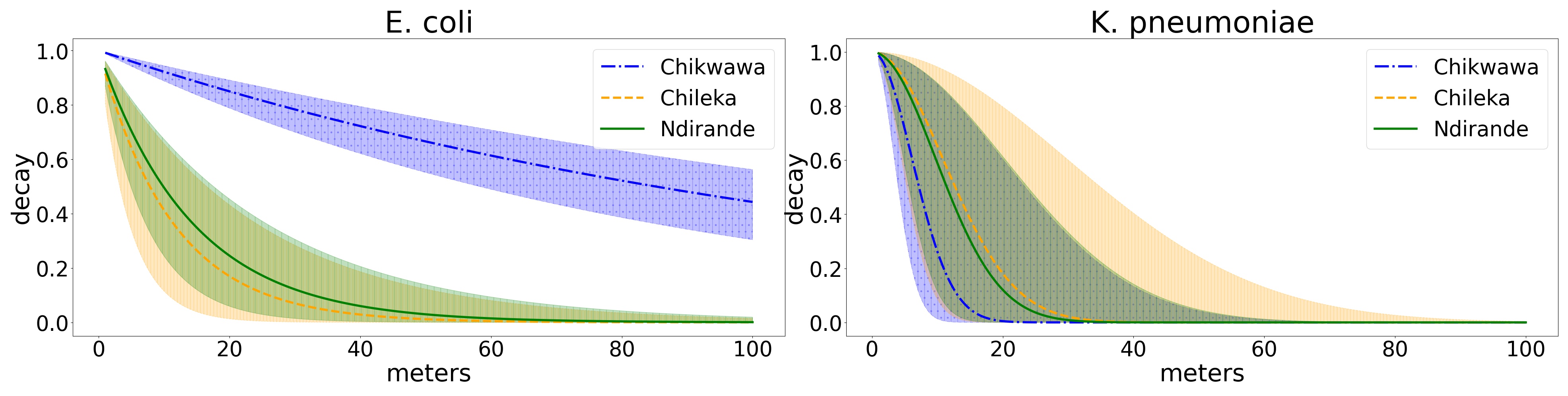}
    \caption{Spatial decay with distance (in metres). On the left \emph{E. coli}, on the right \emph{K. pneumoniae}. Different colours and lines' shapes show different areas. $90 \%$ credible intervals are reported in shaded regions, while lines show the medians.}
    \label{fig:spatial_decay}
\end{figure}

Another interesting aspect of the study is the comparison of the spatial decay parameter $\phi$, which is shown in Figure \ref{fig:spatial_decay}. When comparing bacterial species, we observe that \emph{K. pneumoniae} has a slow decay in space for closer households (Gaussian decay) compared to \emph{E. coli}, \emph{K. pneumoniae} then decays faster compared to \emph{E. coli} for more distant households. This is most likely due to the different ways in which the bacteria transmit. \emph{E. coli} is frequently linked to the environment, especially faecal extraction by humans and animals, hence an individual is more likely to become colonised if living in a contaminated environment, hence we expect it to be more persistent with distance. Colonisation with \emph{K. pneumoniae} typically occurs after direct contact hence it is restricted to the closest neighbours.  

Given the sparsity of our data, seasonality ($\alpha$) is difficult to identify.  However, confirmed seasonality in other studies motivates its inclusion here \citep{Cocker2022}. Casting this as a model choice problem (Section \ref{sec:model_selection}), we find that setting $\alpha=0.6$ or $\alpha=0.8$ gives the largest marginal likelihood. This supports the existence of a strong seasonal effect on household transmission rate and so a big variation between wet and dry season. 

In this study, we find that $\delta = (0,0,0)$ gives the best marginal likelihood, from which we conclude that age, gender and income do not play an important role in driving transmission, which is also consistent with \cite{sammarroEtAl2022,Cocker2022}. In practice, setting $\delta = (0,0,0)$ implies that $I(\delta)^{(k)}=1$, indicating homogeneous transmission rates within each household, consequently suggesting a greater importance of the spatial interactions over the individuals' covariates. 

To conclude, the fixed effect $\epsilon$ is stronger for \emph{E. coli} than for \emph{K. pneumoniae} (with the exception of Chileka). This is consistent with the archetypal nosocomial nature of \emph{K. pneumoniae}, which spreads mainly through direct person-to-person contact \citep{podschun1998klebsiella} and we expect the population dynamic to prevail, i.e. most of the infections are explained by the interactions within and between households. 

\begin{figure}[httb!]
    \centering
    \includegraphics[width=\textwidth]{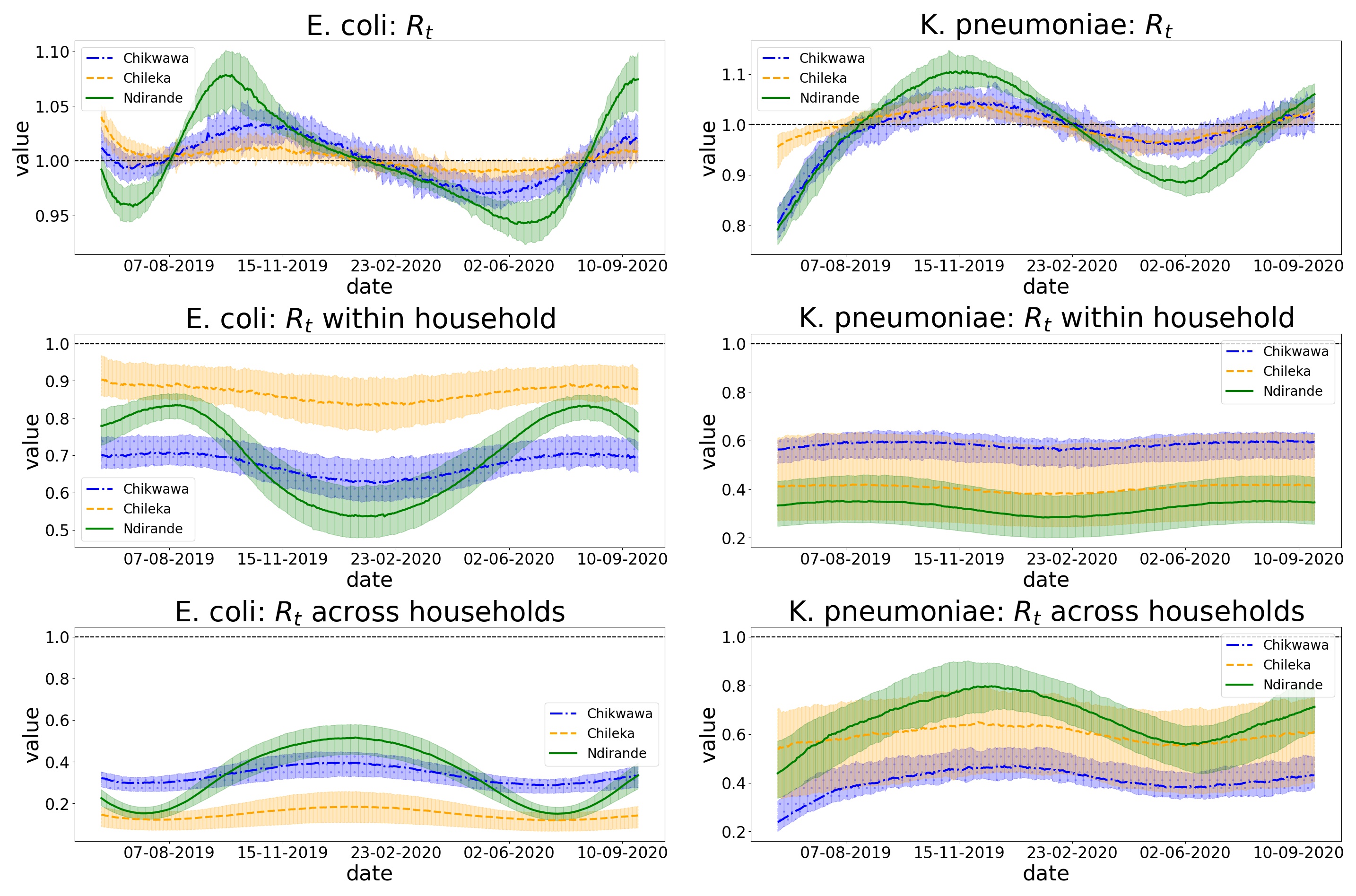}
    \caption{Effective R and its decomposition. \emph{E. coli} is reported in the first column, while \emph{K. pneumoniae} is reported in the second one. The first row shows the effective R, the second row the effective R within household, the third row is the effective R between households. Different colours are associated with different areas. $90 \%$ credible intervals are reported in shaded regions, while lines show the medians.}
    \label{fig:R_t}
\end{figure}

\begin{figure}[b!]
    \centering
    \includegraphics[width=1\textwidth]{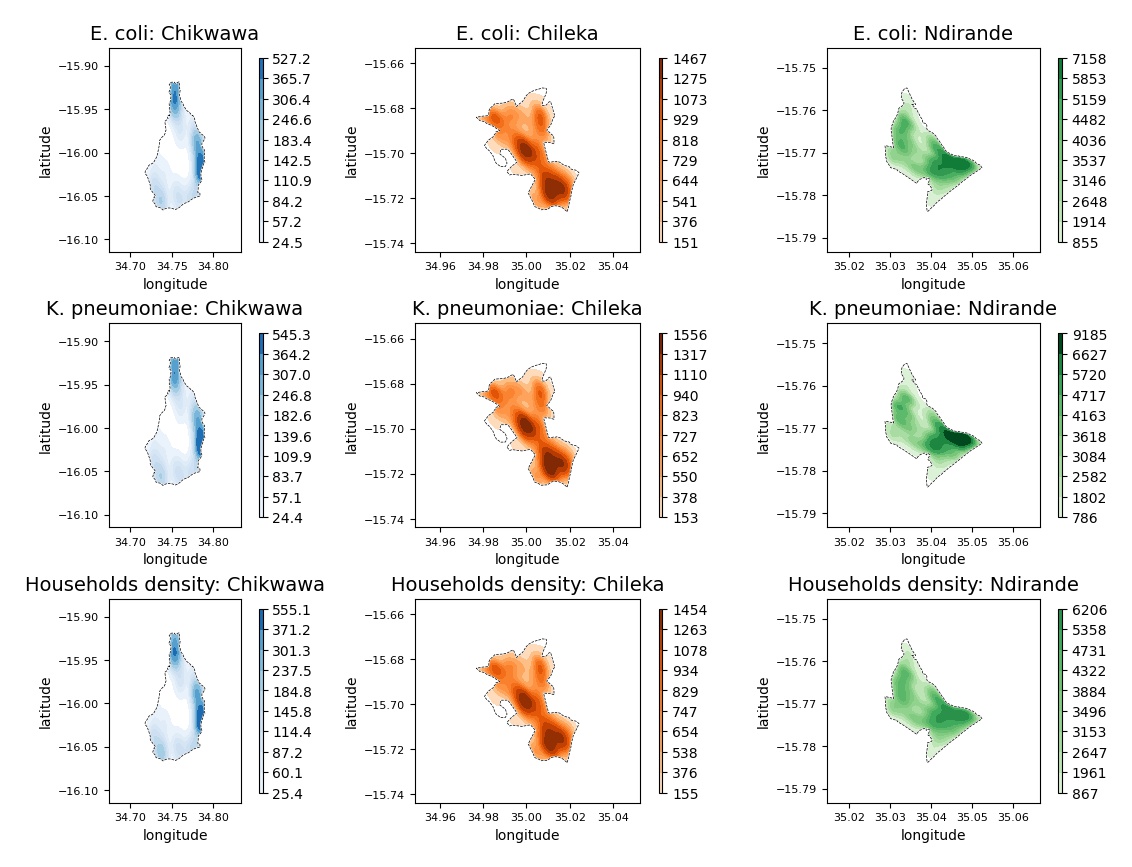}
    \caption{KDE of the spatial density with average colonisation prevalence over time and sampling dimension used as weights. On the columns from left to right: Chikwawa, Chileka, Ndirande. On the rows from top to bottom: \emph{E. coli}, \emph{K. pneumoniae} and the KDE with uniform weights. Different colours are associated with different areas of study and color maps are the same in each area of study.}
    \label{fig:heatmap}
\end{figure}

\subsection{Spatial and temporal incidence}

As already mentioned, SMC$^2$ provides a sample from the posterior distribution over the parameters of interest, which can then be used to sample from the latent process and estimate how colonisation with the bacteria evolved over time and space.

We can measure the colonisation evolution in time with an approximate $R_t$, defined as the expected number of new colonised over the expected number of new uncolonised. This simple approximation also allows to decompose the $R_t$ in an $R_t$ within household and an $R_t$ between households, which includes the fixed effect $\epsilon$. The approximate effective R is reported in Figure \ref{fig:R_t}. We can observe that the effective R is fluctuating above and below 1, showing peaks during the wet season for both \emph{E. coli} and \emph{K. pneumoniae}. We notice a strong within household effective R for \emph{E. coli}, which might indicate inadequate hygiene practices within the household. For \emph{K. pneumoniae}, the between households effective R seems higher than the within household one, suggesting the interaction between households to be the highest source of colonisation, probably due to frequent interaction with neighbours and lack of social distancing.

We can estimate the spatial density of the colonisation, by running a 2 dimensional Kernel density estimation (KDE) on the households and weighting each household with the corresponding average prevalence over time and sampling dimension. Figure \ref{fig:heatmap} shows the KDE estimates for average prevalence from \emph{E. coli}, average prevalence from \emph{K. pneumoniae} and uniform weighting (KDE estimate of the households' density). Comparing with uniform weighting allows us to understand if the spread of the bacteria is uniform in space or if it is particularly concentrated in certain areas. In Chikwawa we observe that the density of \emph{E. coli} is similar to the households' density, hence the bacteria spreads uniformly in space, while \emph{K. pneumoniae} looks particularly intense in the east of the area. For Chileka both bacteria's densities are close to the households' density, hence it seems that they spread uniformly in space. In Ndirande \emph{K. pneumoniae} has a strong prevalence in the south-east of the area, while \emph{E. coli} looks uniform in space.


\section{Conclusion}

We propose to model the spread of AMR bacteria with a partially observed SIS model, called the UC model, where the transmission rate takes into account: within household contacts, between households contacts and spatial decay, seasonality, individuals' covariates and environmental effect. We infer the parameters with the algorithm SMC$^2$, which also allows to perform model selection according to the marginal likelihood on the data. The method is not case specific and can be applied to any AMR bacteria dataset with spatial correlation. We present data on colonisation with ESBL-producing \emph{Escherichia coli} and ESBL-producing \emph{Klebsiella pneumoniae} from three areas in Malawi: Chikwawa, Chileka, Ndirande. As a first step we impute missing data, by following previous studies \citep{darton2017strataa} and then we apply our method to obtain a sample from the posterior distribution over the parameters of interest. From the study, we find \emph{E. coli} to be more persistent in the environment (fixed effect) compared to \emph{K. pneumoniae}, which is in concordance with our knowledge of the bacteria \citep{sammarroEtAl2022,Cocker2022}. We find that setting a high seasonal effect gives higher marginal likelihoods than smaller values, suggesting significant changes in transmission dynamics throughout the year. The effective R helps quantifying the contributions of the within and between households contacts. We also argue that individuals' covariates are not influential in the colonisation process, or at least that our findings prefer models with transmission rates being homogeneous within the households. We also detect geographical hot-spots in the area of Chikwawa for \emph{E. coli} and in the area of Ndirande for \emph{K. pneumoniae}.

There are multiple appealing aspects of this approach. Posterior sampling in epidemiological modelling is a difficult task and it becomes even more challenging when dealing with sparse data. Our method provides an efficient way of performing Bayesian inference on the parameters of a SIS model that is both spatially and temporally sparse. Moreover, it is accompanied by a principled way of performing model selection and supported by strong mathematical results \citep{chopin2013smc2}. However, the pivotal point of our method is the interpretation, all the parameters and structures in the model have a direct connection with real-world data and it is particularly reassuring that our experimental results agree with the scientific knowledge that we have on the considered bacterial species \citep{sammarroEtAl2022,Cocker2022}. In addition, our approach provides simple ways of building useful tools for investigating outbreaks and tailored public health interventions to contain pathogens.

To conclude, there are several strands of research that might follow from this work. There are lots of technical questions related to SMC$^2$ and what are the best ways of choosing: tuning parameters, proposal distribution, etc. The procedure has potential utility outside the field of epidemiology to any dataset with spatial interactions, however the computational cost may become prohibitive. Finally, this study is restricted to an SIS model, however adapting it to more complex compartmental models is straightforward and the same methodology can be applied to any epidemiological model. 

\section*{Acknowledgements}

This work is supported by  MR/S004793/1 (Drivers of Resistance in Uganda and Malawi: The DRUM Consortium), EPSRC grants EP/R018561/1 (Bayes4Health) and EP/R034710/1 (CoSInES).

\bibliographystyle{chicago}
\bibliography{references.bib}

\appendix

\section{Data}

\subsection{ESBL \emph{E. coli} and \emph{K. pneumoniae} in Malawi}

In Malawi, where the incidence of severe bacterial infection is high, inadequate water, sanitation and hygiene infrastructure combined with a poorly regulated antimicrobial market have made the country an ideal place for antimicrobial resistance (AMR) to transmit and persist \citep{chikowe2018amoxicillin}. More and more infections become locally untreatable, due to the rapid emergence of resistant bacteria such as Extended-Spectrum $\beta$-Lactamase(ESBL)-producing \emph{E. coli} and \emph{K. pneumoniae}. These bacterial species have become resistant to 3rd-generation cephalosporin, the first and last antimicrobial of choice in much of sub-Saharan Africa \citep{cocker2022drivers}. In fact, ESBL resistance in both \emph{E. coli} and \emph{K. pneumoniae} has significantly increased in the last two decades \citep{musicha2017trends}. 

The Drivers of resistance in Uganda and Malawi (DRUM) consortium is a trans-disciplinary collaboration, aiming to study AMR transmission across urban and rural sites in a One Health setting: areas with different human and animal population densities, and different levels of affluence and infrastructure \citep{cocker2022drivers}. 

The samples were collected over a time span of about 1 year and 5 months (from 29-04-2019 to 24-09-2020) covering both the wet (November-April) and dry (May-October) seasons in three study areas in Malawi: Chikwawa, Chileka and Ndirande. The dataset consists of positive-negative sample results for colonisation with ESBL-producing \emph{E. coli} and
ESBL-producing \emph{K. pneumoniae}, individual ID, household ID, household location, individual and individual-level variables: gender, income and age, extracted from the complete dataset. The considered data contain 1492 samples, from 373 individuals in 129 different households. The distribution of our samples per study area is detailed in Table \ref{hip}.

\begin{table}
\centering
\caption{\label{hip}Distribution of households, individuals and samples per study area} 
\fbox{%
\begin{tabular}{c|c|c|c}
\bf  & \bf Ndirande & \bf Chikwawa & \bf Chileka  \\
\hline
Households & 37 & 55 & 37   \\
\hline
Individuals & 96 & 160 & 117   \\
\hline
Samples & 384 & 640 & 468 \\
\end {tabular}
}
\end{table}

\begin{figure}[ht]
\centering
\includegraphics[scale=0.6]{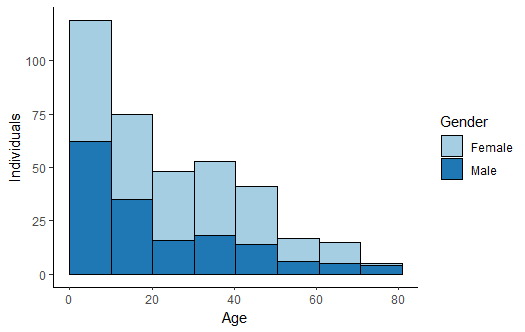}
\captionsetup{justification=centering}
 \caption{Distribution of age and gender for the individuals}
  \label{agesex_desc}
\end{figure}

As Figure \ref{agesex_desc} shows, 52\% of the individuals were under the age of 20 years old, 38.1\% were between 21 and 50 years old, 8.6\% were between 51 and 70 years old and the last 1.3\% were over 70 years old. Over all the individuals, 57.1\% were female with slightly varying proportions in each age group.

\begin{figure}[ht]
\centering
 \includegraphics[scale=0.6]{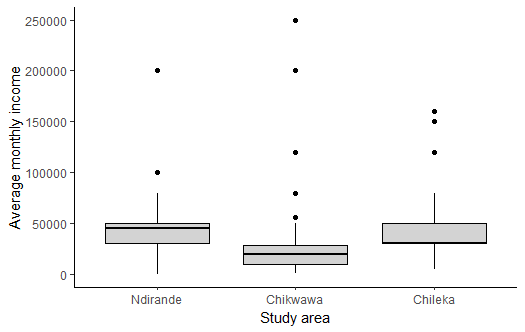}
\captionsetup{justification=centering}
 \caption{Distribution of household monthly income in each study area}
  \label{income_desc}
\end{figure}

The boxplots in Figure \ref{income_desc} indicated a slight difference in the distribution of income among the different study areas. Chikwawa was the area with the lowest monthly income with a median income of only 20000 mwk compared to 30000 mwk in Chileka and close to 45000 mwk in Ndirande. There appeared to be less variation in the monthly income within the Chikwawa study area with a range of around 50000 mwk. In comparison, the ranges of Chileka and Ndirande were almost close to 80000 mwk. Each area appeared to have a couple of outlier houses with a higher income than average.

Overall, the prevalence of ESBL-producing \emph{E. coli} in our samples was 38.3\% and the prevalence of ESBL-producing \emph{K. pneumoniae} was 11.7\%. At the first visit, 141 were positive for ESBL-producing \emph{E. coli} (37.8\%) and 45 were positive for ESBL-producing \emph{K. pneumoniae} (12.1\%). At the second visit, 133 were positive for ESBL-producing \emph{E. coli} (35.7\%) and 43 were positive for ESBL-producing \emph{K. pneumoniae} (11.5\%). At the third visit, 145 were positive for ESBL-producing \emph{E. coli} (38.9\%) and 32 were positive for ESBL-producing \emph{K. pneumoniae} (8.6\%). At the last visit, 153 were positive for ESBL-producing \emph{E. coli} (41.0\%) and 55 were positive for ESBL-producing \emph{K. pneumoniae} (14.8\%). The prevalences can be found in Table \ref{prevtable}.

\begin{table}[h!]
\caption{\label{prevtable}Prevalence of ESBL-producing \emph{E. coli} (Ec) and ESBL-producing \emph{K. pneumoniae} (K)}
\centering
\fbox{
\begin{tabular}{c|c|c|c|c}
Visit & ESBL & Positive (P) & Total (T) & (P/T)*100  \\
\hline
\multirow{2}{*}{1} & Ec &   141     &   373   & 37.8\%  \\
&  K & 45  &  373  & 12.1\%\\[1ex]
\hline
\multirow{2}{*}{2} & Ec &     133       &    373    & 35.7\%   \\ 
& K &   43    & 373  & 11.5\% \\
\hline
\multirow{2}{*}{3} & Ec &145   &    373    &38.9\%   \\ 
& K &  32  &   373    & 8.6\% \\
\hline
\multirow{2}{*}{4} &  Ec & 153  & 373  & 41.0\%  \\
 &  K &  55  &   373   & 14.8\% \\
\end{tabular}
}
\end{table}

\subsection{Simulation of synthetic population}

In order to simulate a ``large'' community into which our data sample can be embedded, we use a mechanistic resampling approach.  Let $\mathcal{H}_S \subset \mathcal{H}$ represent our panel of households recruited into the DRUM study, assumed to be an unbiased sample from a population of size $n_h$ households. Additionally, let $L_{\mathcal{H}}$ be a set of household locations within the study area identified using OpenStreetMap.

To construct the synthetic population, a household in $\mathcal{H}_S$ is sampled with replacement and randomly allocated to a household location sampled without replacement from $L_{\mathcal{H}}$.  This process is repeated as many times as is required such that the total number of individuals represented by the sampled collection of households reaches the simulated population size $n_I$.  

A peculiarity of the DRUM data is that the complete age structure of a household is known only if \emph{all} individuals within the household were sampled.  For households where a fraction of the inhabitants were sampled, ages were only known for the individuals who had been sampled albeit with the total household size known.  For these households, the age-structure of the unsampled individuals was simulated by ``borrowing'' the age-structure from the most closely matched fully-sampled household.

An instance of synthetic data generated with the algorithm is available at \href{https://github.com/LorenzoRimella/SMC2-ILM}{https://github.com/LorenzoRimella/SMC2-ILM} in the "Cleaning/Data/Synthetic" folder.

More details on how to generate the synthetic population can be found at: \\ \href{https://zenodo.org/record/7007232#.Yv5EZS6SmUk}{https://zenodo.org/record/7007232\#.Yv5EZS6SmUk}.


\section{Model}
\subsection{Agent-based UC model} \label{sec:appUCmodel}
The individual-based model used in the SMC$^2$ consider an $h>0$ Euler discretization of the corresponding continuous in time process:
\begin{equation}\label{eq:dynamic_h_appendix}
    C_0^{(k)} \sim \mathcal{B}(1-e^{-\lambda_0}), \quad C_{t+h}^{(k)} \sim 
    \begin{cases}
        \mathcal{B}\left(1-e^{-h{\lambda}_t(\theta)^{(k)}}\right) & \text{if } C_{t}^{(k)}=0\\
        \mathcal{B}(e^{-h\gamma})              & \text{if } C_{t}^{(k)}=1\\
    \end{cases}.
\end{equation}
This representation allows to model coarser discretization (e.g. weekly discretization when $h=7$), which can be used when there is a significant time sparsity in the data. Obviously a bigger $h$ implies a worse approximation of the corresponding continuous dynamic, with the extreme case being all the U becoming C and all the C becoming U. At the same time a too small $h$ might result in redundant dynamics, it is likely to stay in the same state, and high randomness, when changing state the epidemic trajectories look very different. For our UC model we can easily identify the threshold $1 \slash \gamma$, which is the mean recovery time of an infected individual. Choosing $h > 1 \slash \gamma$ would cause all the infected to recover at the next simulation and we might even lose some of the infection events, a sensible choice is then $h \leq 1 \slash \gamma$. For our experiment we find $h=7$ to perform significantly better than other choices.


\section{Inference}

\subsection{Additional information on SMC$^2$}
SMC$^2$ \citep{chopin2013smc2} approximates the posterior distribution over $\theta, C_{0}, \dots, C_t$ on the same vein of particle filter, precisely it combines IBIS \citep{chopin2004central} with particle filter and it intuitively creates an SMC for the parameters and an SMC for the latent process $(C_t)_{t \geq 0}$. The algorithm requires the same input of IBIS plus a particle filter with the corresponding particle filter step. We report our version of SMC$^2$ in the main paper and we specify here some additional details on the implementation. 

\paragraph{Number of particles per area of study}
For our experiments we consider an SMC$^2$ with $P_\theta=300,200,150$ for Chikwawa, Chileka and Ndirande and $P=300,200,150$ for Chikwawa, Chileka and Ndirande in both bacteria.

\paragraph{Parameters priors}
We use the following priors for the parameters:
\begin{itemize}
    \item $\log(\beta_1) \sim \mathcal{N}(-3,1)$;
    \item $\log(\beta_2) \sim \mathcal{N}(-3,1)$;
    \item $\log(\phi) \sim \mathcal{N}(-3,1)$;
    \item $\log(\delta) \sim \mathcal{N}((0,0,0), diag((0.5,0.5,0.5)))$;
    \item $\log(\alpha) \sim \mathcal{B}eta(50,50)$;
    \item $\log(\epsilon) \sim \mathcal{N}(-5,1)$;
\end{itemize}
where $\mathcal{N}(\cdot,\cdot)$ is a Gaussian distribution (multivariate if using a vector and a matrix as parameters), $diag(\cdot)$ is the diagonal matrix with diagonal $\cdot$ and $\mathcal{B}eta(\cdot,\cdot)$ is a Beta distribution. The priors are then combine in a product to define the overall prior over $\theta$. Remark that in some of our experimental settings $\alpha, \delta$ are fixed, in that case the priors are simply removed from the product.

\paragraph{Parameters proposal}
Given that we have a full sample $(\theta^{(1)}, \dots, \theta^{(N_\theta)})$ we can build an adaptive proposal by computing:
\begin{align*}
    \hat{\mu} = \sum_{m=1}^{P_\theta} w_\theta^m \theta^m, \quad
    \hat{\Sigma} = \sum_{m=1}^{P_\theta}  w_\theta^m \left ( \theta^m - \hat{\mu} \right )\left ( \theta^m - \hat{\mu} \right )^{\mathrm{T}},
\end{align*}
and so set $p(\tilde{\theta}|\theta)$ to be either $\mathcal{N}(\theta, \hat{\Sigma})$ or $\mathcal{N}(\hat{\mu}, \hat{\Sigma})$. The covariance matrix might be tricky to set, indeed it frequent to get a covariance matrix that is not positive definite. An easy fix is to set $\hat{\Sigma} = \hat{\Sigma}+b I$ with $I$ identity matrix. Sometimes it is also useful to increase or decrease the $\hat{\Sigma}$ to allow further or closer jumps, this can be easily done by setting $\hat{\Sigma} =  c\hat{\Sigma}$. In our experiments, we consider an adaptive proposal $\mathcal{N}(\hat{\mu}, c\hat{\Sigma}+b I)$ with $b=0.25$ and $c=0.001$ to ensure an acceptance rate of about $23 \%$ in all the experiments. 

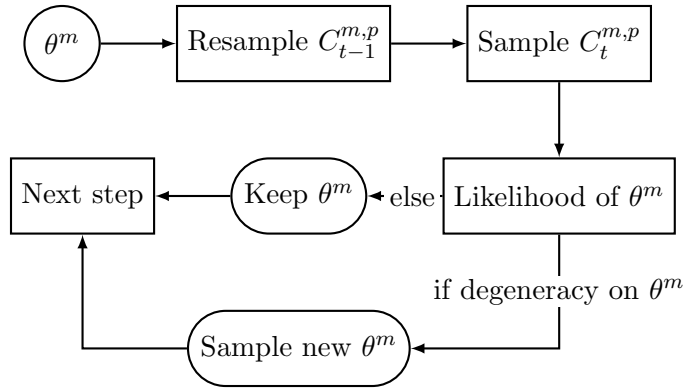
\begin{figure}[httb!]
    \centering
    \begin{tikzpicture}[font=\small,thick]
        \node[draw, rounded rectangle, minimum width  = 1cm, minimum height = 1cm] (theta) {$\theta^{m}$};
        \node[draw, right=of theta, rectangle, minimum width  = 1cm, minimum height = 1cm] (C_tm1) {Resample $C_{t-1}^{m,p}$};
        \node[draw, right=of C_tm1, rectangle, minimum width  = 1cm, minimum height = 1cm] (C_t) {Sample $C_t^{m,p}$};
        \node[draw, below=of C_t, rectangle, minimum width  = 1cm, minimum height = 1cm] (likelihood) {Likelihood of $\theta^{m}$};
        \node[draw, left=of likelihood, rounded rectangle, minimum width  = 1cm, minimum height = 1cm] (oldtheta) {Keep $\theta^{m}$};
        \node[draw, below=of oldtheta, rounded rectangle, minimum width  = 1cm, minimum height = 1cm] (newtheta) {Sample new $\theta^m$};
        \node[draw, left=of oldtheta, rectangle, minimum width  = 1cm, minimum height = 1cm] (start) {Next step};
        
        \draw[-latex] (theta) edge (C_tm1);
        \draw[-latex] (C_tm1) edge (C_t);
        \draw[-latex] (C_t) edge (likelihood);
        \draw[-latex] (likelihood) edge node[pos=0.4,fill=white,inner sep=2pt]{else}(oldtheta) 
                      (likelihood) |- (newtheta) node[pos=0.25,fill=white,inner sep=1pt]{if degeneracy on $\theta^{m}$};
        \draw[-latex] (oldtheta) edge (start);
        \draw[-latex] (newtheta) -| (start);
    \end{tikzpicture}
    \caption{The algorithm flow of a general $t$ step of SMC$^2$.}
    \label{fig:diag_SMC2}
\end{figure}

\paragraph{Rejuvenation step}
In SMC$^2$ a key step is the rejuvenation step (from line 16 onwards), here each of the current combination of parameters is evaluated and eventually replaced if needed. The rejuvenation step is run only if a degeneracy criteria on $w_\theta^m$ is satisfy, for our implementation we follow the routine suggested by \cite{chopin2013smc2} and we run a rejuvenation step whenever the ESS $[\sum_{m=1}^{P_\theta} (w_\theta^m)^2]\slash (\sum_{m=1}^{P_\theta} w_\theta^m)^2$ is below $P_\theta/2$. We implement the rejuvenation step as follows:
\begin{itemize}
    \item[RESAMPLE:] we resample $\theta^{m}$ proportionally to $w_\theta^m$:
    \begin{align*}
        \theta^{m} \sim \sum_{\bar{m}=1}^{P_\theta} w^{\bar{m}}_\theta \delta_{\theta^{\bar{m}}}\left (\cdot \right);
    \end{align*}
    \item[MUTATION:] we propose new $\tilde{\theta}^{m}$ according to our parameters proposal:
    $$
    \tilde{\theta}^{m} \sim p\left (\tilde{\theta}^{m}|\theta^{m}\right),
    $$
    we run an APF for each new $\tilde{\theta}^{m}$ and get new particles for the latent process $C_t^{m,p}$ and a particle approximation of the likelihood $\mathcal{L}_t(\tilde{\theta}^m)$;
    \item[REPLACE:] run a Metropolis-Hastings step using the approximate likelihood:
    $$
    \left ({\theta}^{m}, {C}_t^{m,p}, {\mathcal{L}}_t\left ({\theta}^{m}\right) \right ) \leftarrow
        \begin{cases}
            \left (\tilde{\theta}^{m}, \tilde{C}_t^{m,p}, {\mathcal{L}}_t\left (\tilde{\theta}^{m}\right) \right ) & \text{w.p. } \frac{p\left (\tilde{\theta}^{m}\right)p\left ({\theta}^{m}|\tilde{\theta}^{m}\right)}{p\left ({\theta}^{m}\right)p\left (\tilde{\theta}^{m}|{\theta}^{m}\right)} \frac{{\mathcal{L}}_t\left (\tilde{\theta}^{m}\right)}{{\mathcal{L}}_t\left ({\theta}^{m}\right)}\\
            \left ({\theta}^{m}, {C}_t^{m,p}, {\mathcal{L}}_t\left ({\theta}^{m}\right) \right ) & \text{otherwise}
        \end{cases}.
    $$
    and set $w_\theta^m \leftarrow 1$.
\end{itemize}
Note that the rejuvenation step requires to run a full particle filter from scratch with the proposed parameters, which is what Particle Marginal Metropolis-Hastings \citep{andrieu2010particle} does on the full path. However, there is a dynamic learning of the parameters, indeed PMCMC algorithms run a particle filter on the whole data for a given parameters, while SMC$^2$ requires a particle filter on the data up to the current time step only during the rejuvenation step. Diagrams in Figure \ref{fig:diag_SMC2} shows the flow at time $t$ of SMC$^2$.

As already mentioned, one of the key output of SMC$^2$ is the estimator of the marginal likelihood of the model $\prod_{s=1}^t \hat{p} (y_t|y_{1:t-1})$ where:
\begin{align*}
\hat{p}\left (y_s|y_{1:s-1} \right) \coloneqq
\frac{1}{P P_\theta}\sum_{m=1}^{P_\theta} w^m_\theta  \sum_{p=1}^{P} w_t^{m,p}.
\end{align*}

\paragraph{Comments on APF}
We employ an auxiliary particle filter (APF) as a particle filter routine. APF is convenient for this scenario because we can compute $p(C_t|C_{t-1}, Y_t)$ in closed form:
\begin{equation}
    p(C_t|C_{t-1}, Y_t) = \frac{p(Y_t|C_{t})p(C_t|C_{t-1})}{\sum_{C_t}p(Y_t|C_{t})p(C_t|C_{t-1})}
\end{equation}
where $p(C_t|C_{t-1})$ is the transition kernel as in \eqref{eq:dynamic_h_appendix} with $h=1$ and $p(Y_t|C_{t})$ is the emission distribution:
\begin{equation}
    Y_t^{(k)}|C_t^{(k)} \sim 
    \begin{cases}
        \mathcal{B}(s_e)  & \text{if }k\in \mathrm{R}_t, \text{ } C_t^{(k)}=1\\ 
        \mathcal{B}(1-s_p) & \text{if }k\in \mathrm{R}_t, \text{ } C_t^{(k)}=0\\
        \text{NA} & \text{otherwise}
    \end{cases}.
\end{equation}
Note that the marginalization at the denominator is feasible because both the transition kernel and the emission distribution factorize over the individuals resulting in a computational cost $\mathcal{O}(2 N_I)$ when switching the product and the sum. 

However, it is difficult to deal with the time sparsity, it might be that most of the time observation are not available at time $t$ and we end up proposing with the transition kernel only, which is likely to select bad future particles. As an alternative we can employ the Euler discretization with $h$ to reduce the Monte Carlo error, indeed a bigger $h$ makes the whole procedure less sensible to randomness by reducing the number of proposed particles and eventually include the observation in the proposal (or get closer to the observation). In this case our proposal is going to be $p(C_{t+h}|C_{t}, Y_{t+h})$, with the proposal becoming $p(C_{t+h}|C_{t})$ if $Y_{t+h}=NA$. Observe that we might have non-regular intervals between observation and it might even be that the interval is not divisible by $h$, this problem is solved by defining a schedule of simulations between observations as described in the main paper. 


\section{Experiments}

 \subsection{Model selection}
\begin{figure}[httb!]
     \centering
     \includegraphics[width=0.8\textwidth]{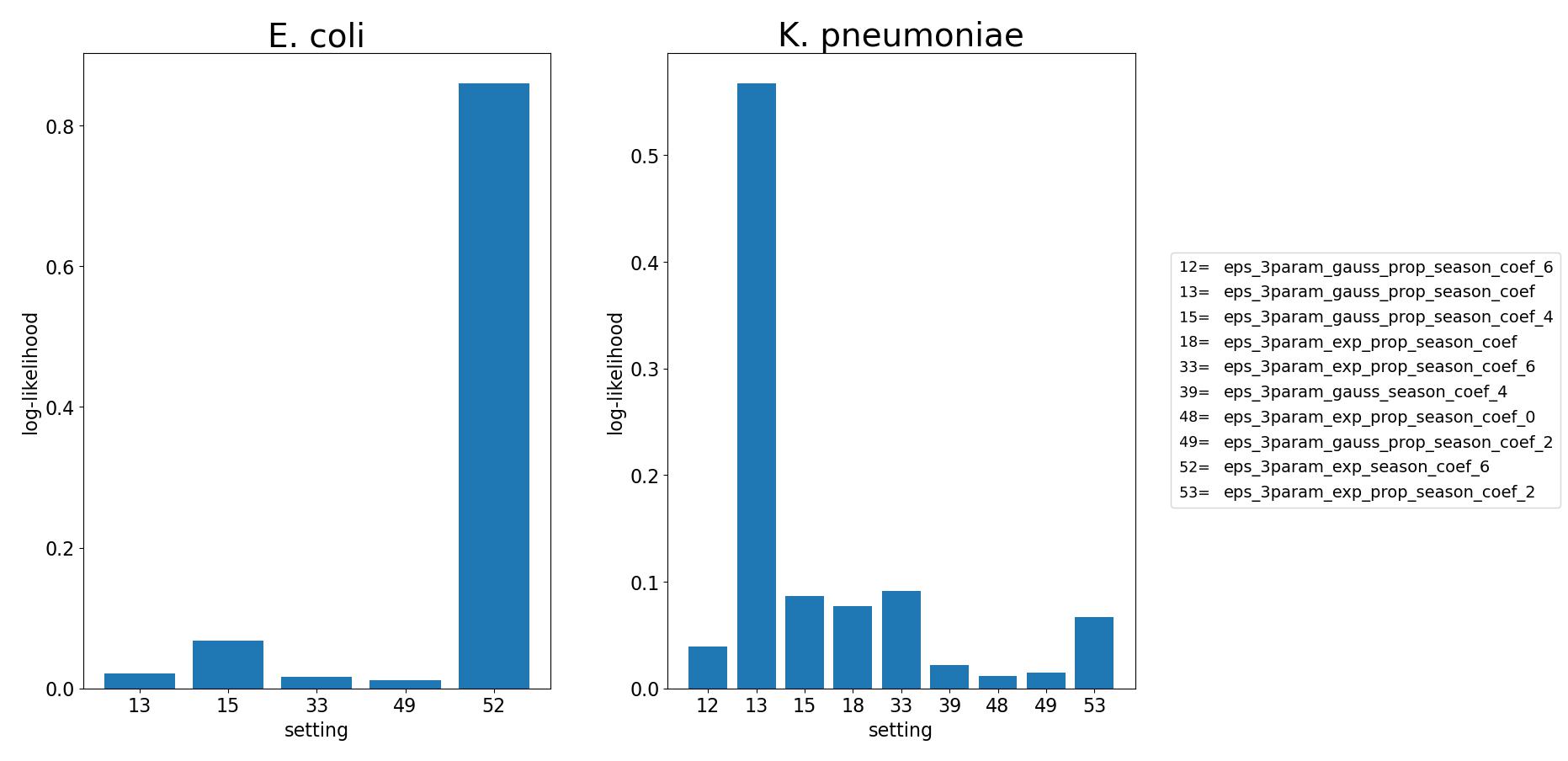}
     \caption{Best models in terms of cumulative marginal likelihood for each bacterium. On the left \emph{E. Coli} on the right \emph{K. Pneumoniae}. Numbers label different models. The legend provides the correspondence of the numbers with their unique identifier.}
     \label{fig:evidence_hist}
 \end{figure}

 SMC$^2$ provides an estimate $\prod_{s=1}^t \hat{p} (y_s|y_{1:s-1})$ for the marginal likelihood, which can be used for model selection. As mentioned in the main paper we try 55 different models and select the best one in terms of best marginal likelihood under uniform prior. To do so we cumulate the marginal likelihood over the areas of study and we put a uniform prior over the 55 models. Figure \ref{fig:evidence_hist} reports the best models for each bacterium. We used unique strings as identifier of the 55 models. Each string is formed by combining sub-strings which have different meaning. For the strings in the legend of Figure \ref{fig:evidence_hist} we have:
 \begin{itemize}
     \item ``\textit{eps}'': $\epsilon$ is learned under a Gaussian prior;
     \item ``\textit{3param}'': we learn $\beta_1,\beta_2, \phi$;
     \item ``\textit{exp}'': we use an exponential spatial decay;
     \item ``\textit{gauss}'': we use a Gaussian spatial decay;
     \item ``\textit{prop}'': we use $\kappa_1(H) = |H|$ (if not present we use $\kappa_1(H) = 1$);
     \item ``\textit{season\_coef\_6}'': we fix $\alpha = 0.6$ (similar for ``$\_2$'', ``$\_4$'' and if not present we used $\alpha = 0.8$). 
 \end{itemize}

  \begin{figure}[httb!]
     \centering
     \includegraphics[width=0.8\textwidth]{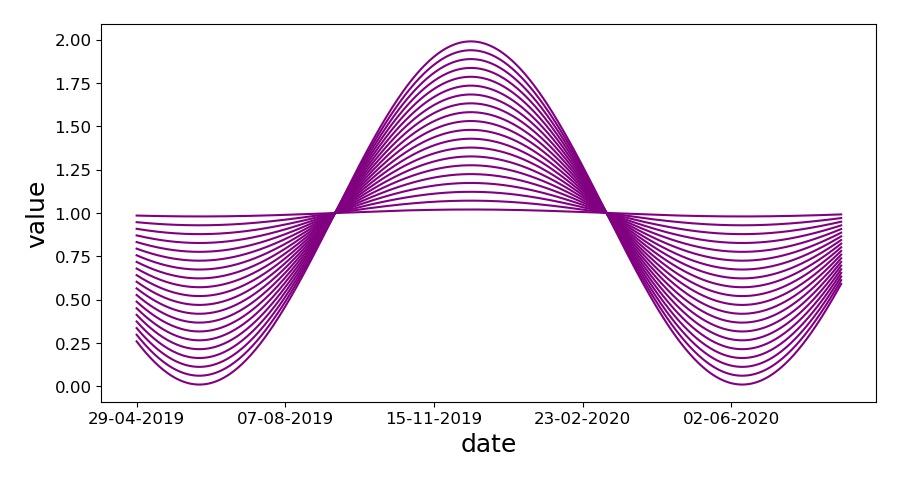}
     \caption{Seasonal coefficient when varying $\alpha$.}
     \label{fig:change_seasonality}
 \end{figure}
 
 \begin{figure}[httb!]
 \centering
 \includegraphics[width=0.8\textwidth]{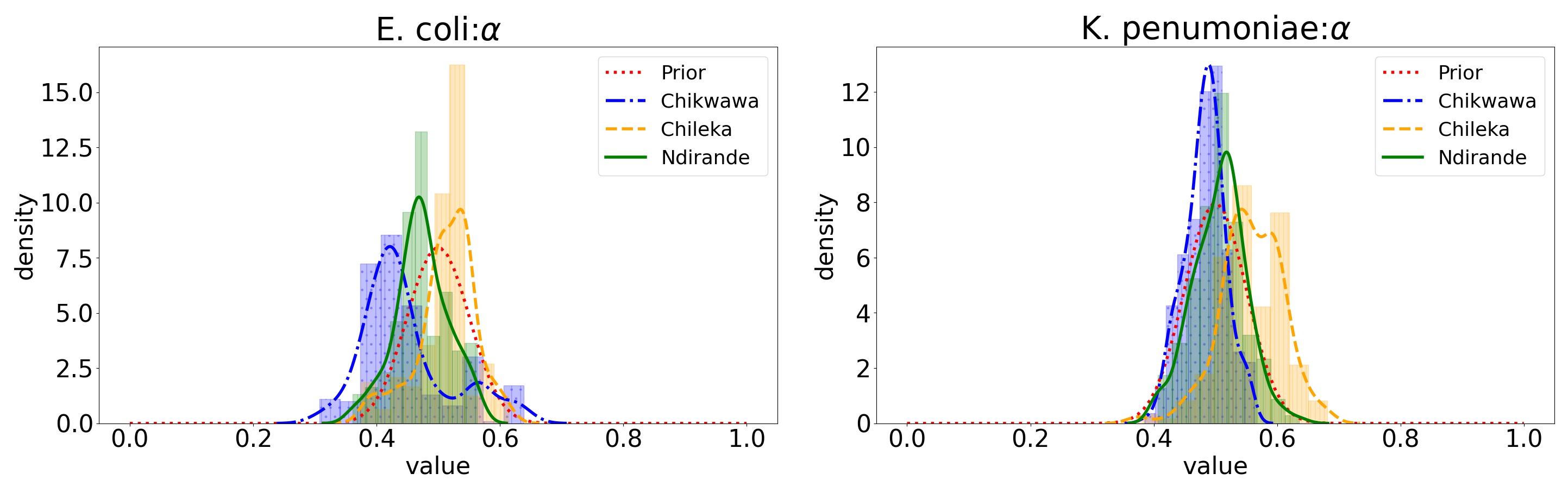}
 \caption{Posterior distribution over $\alpha$.}
 \label{fig:alpha_posterior}
\end{figure}
 
 \subsection{Seasonality}
\begin{figure}[httb!]
 \centering
 \includegraphics[width=0.7\textwidth]{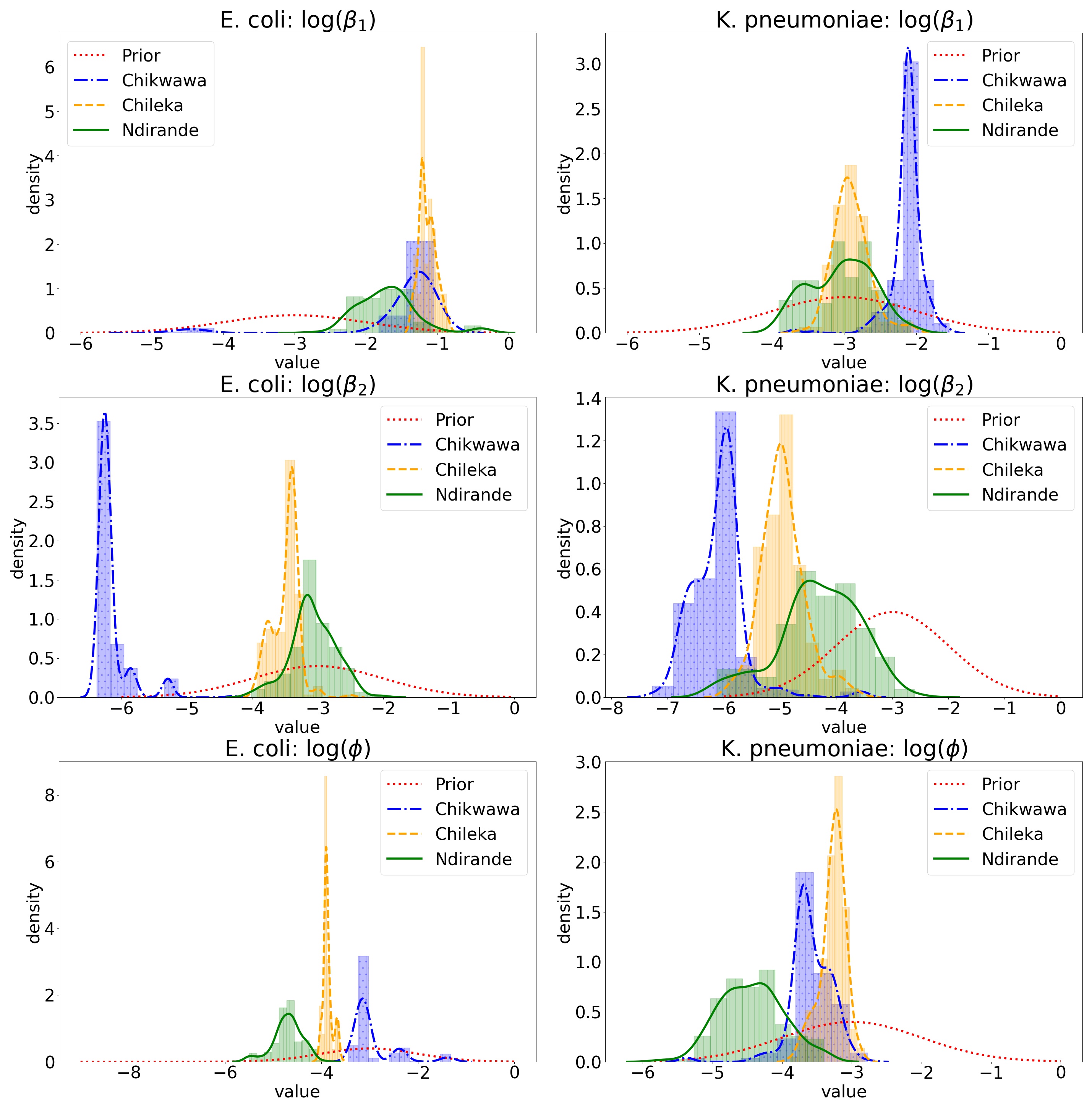}
 \caption{Posterior distribution over $\beta_1,\beta_2,\phi$ when learning $\alpha$.}
 \label{fig:multimod_posterior}
\end{figure}

 As already mentioned in the main paper, the seasonality is a key part of our model. We are assuming a different spread of the bacteria during different seasons and we show the value of the seasonal coefficient over time when varying $\alpha$ in Figure \ref{fig:change_seasonality}. In terms of inference, we initially try to learn $\alpha$ with a $\mathcal{B}eta(50,50)$ prior and we find posterior distributions ranging around the interval $(0.4,0.6)$ as in Figure \ref{fig:alpha_posterior}. However, we find learning $\alpha$ to cause identifiability issues in the other parameters, see multimodality in Figure \ref{fig:multimod_posterior}. Hence, guided by the initial findings on the posterior distribution over $\alpha$, we choose the grid $(0.2,0.4,0.6,0.8)$ and learn $\alpha$ in this grid. This results in improved identifiability of the model and also higher marginal likelihood. Over the grid we find $0.6,0.8$ to be the values associated to the highest marginal likelihood, suggesting a strong seasonal effect.

  \begin{figure}[httb!]
     \centering
     \includegraphics[width=\textwidth]{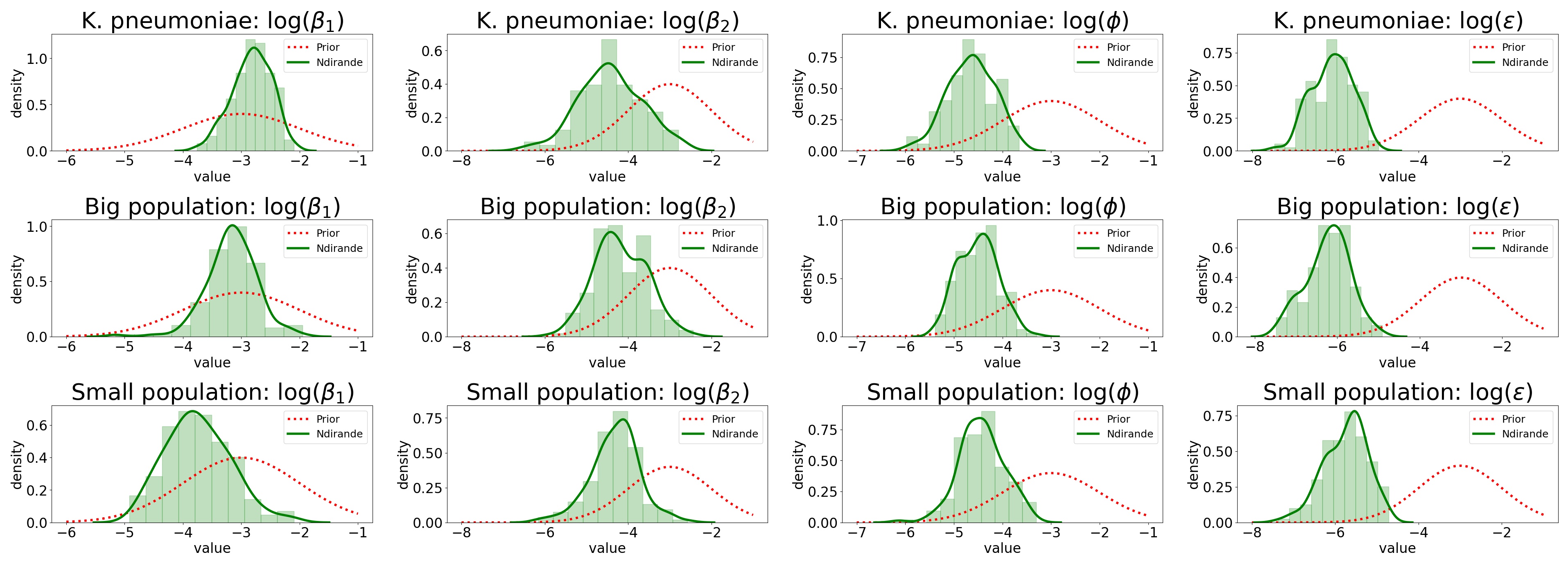}
     \caption{Posterior distribution for Ndirande for the population from the experiments in the main paper, for a bigger population and a smaller population. On the columns from left to right: posterior distribution for $\beta_1,\beta_2,\phi,\epsilon$. On the rows from top to bottom: population from the main paper, big population and small population.}
     \label{fig:invariance}
\end{figure}
 
\subsection{Scale invariance}
We create a synthetic population of 36314 individuals for Ndirande, 13337 individuals for Chileka and 9678 individuals for Chikwawa which are smaller than the actual population in the areas of study. Fitting on a smaller population is a forced choice when dealing with high-dimensional problem and we try to optimize both population size and particle size. However, we argue that we are above the invariance property threshold, i.e. above a certain population size inference for epidemiological models do not change. We check the invariance property empirically by generating $2$ new synthetic populations for Ndirande: a small population of 25312 individuals in 5522 households, and a big population of 47009 individuals in 10219 households. We run the SMC$^2$ algorithm on \emph{K. pneumoniae} with $P_\theta=150,P=150$ for both the new populations and old population and we infer $\beta_1,\beta_2,\phi,\epsilon$ while keeping $\kappa_1(H)=|H|,\kappa_2(H)=|H|, \delta=(0,0,0), \alpha = 0.8$ and $f_\phi$ being Gaussian. Posteriors distribution are reported in Figure \ref{fig:invariance}.  We can observe that the posterior distribution for our population and for the bigger population are centered in the same regions, while smaller population seems to slightly underestimate $\beta_1$. This is due to the scale invariance property of epidemics models. Indeed, we expect that over a certain population threshold increasing the population size $n_I$ more does not change the dynamic of the infections. One can then consider small, but big enough, population sizes to infer the parameter of a bigger population, with the advantage of reducing the computational cost and the variance of the particle approximations. 

\end{document}